\documentclass[mreferee,timet]{gji}
\usepackage{footnote}
\usepackage{algorithm}
\usepackage{algorithmic}
\usepackage[pdftex]{graphicx}
\usepackage{epstopdf}
\usepackage{epsfig,subfigure}
\usepackage{epsf}
\usepackage{graphics}
\usepackage{url}
\usepackage{color}
\usepackage{amssymb}
\usepackage{amsmath}
\usepackage{cleveref}

\graphicspath{{./Figures/}}

\addtocounter{footnote}{0}
\newcommand{\bfm}{\mathbf{m}}
\newcommand{\bfd}{\mathbf{d}}
\newcommand{\bfl}{\boldsymbol{\ell}}
\newcommand{\bfx}{\mathbf{x}}

\newcommand{\bfp}{\mathbf{p}}
\newcommand{\bfr}{\mathbf{r}}
\newcommand{\bfy}{\mathbf{y}}

\newcommand{\bfv}{\mathbf{v}}

\newcommand{\WL}{W_{{L}_1}}

\newcommand{\bfalpha}{\boldsymbol{\alpha}}

\newcommand{\bfdo}{\mathbf{d}_{\mathrm{obs}}}

\newcommand{\bfde}{\mathbf{d}_{\mathrm{exact}}}
\newcommand{\bfma}{\mathbf{m}_{\mathrm{apr}}}
\newcommand{\bfme}{\mathbf{m}_{\mathrm{exact}}}

\newcommand{\Wd}{W_{\bfd}}

\newcommand{\rtilde}{\tilde{ \bfr}}
\newcommand{\bfw}{\mathbf{w}}

\newcommand{\At}{\tilde{A}}
\newcommand{\mt}{\tilde{\bfm}}
\newcommand{\mat}{\tilde{\mathbf{m}}_{\mathrm{apr}}}
\newcommand{\dotilde}{\tilde{\mathbf{d}}_{\mathrm{obs}}}

\newcommand{\bfzero}{\mathbf{0}}

\newcommand{\bfeye}{\mathbf{I}}

\newcommand{\Wdepth}{W_{\mathrm{depth}}}
\newcommand{\Wh}{W_{\mathrm{hard}}}

\newcommand{\Rn}{\mathcal{R}^{n}}
\newcommand{\Rm}{\mathcal{R}^{m}}
\newcommand{\Rnn}{\mathcal{R}^{n \times n}}

\newcommand{\Rsxy}{\mathcal{R}^{(s_x+n_x-1) \times (s_y+n_y-1)}}

\newcommand{\twoDFFT}{\texttt{2DFFT}}
\newcommand{\bttb}{\texttt{BTTB}}
\newcommand{\bccb}{\texttt{BCCB}}
\newcommand{\rrcg}{\texttt{RRCG}}
\newcommand{\cg}{\texttt{CG}}
\newcommand{\ffttwo}{\texttt{FFT2}}
\newcommand{\iffttwo}{\texttt{IFFT2}}
\title[Large-scale joint inversion with Gramian constraint]{Large-scale focusing joint inversion of gravity and magnetic data with Gramian constraint}
\author[S. Vatankhah, R. A. Renaut, X. Huang, K. Mickus, M. Gharloghi]{Saeed Vatankhah $^1$$^,$$^2$, Rosemary A. Renaut $^3$, Xingguo Huang $^1$, \\\LARGE{\textup{Kevin Mickus $^4$,
and Mostafa Gharloghi $^2$}} \\
$^1$ College of Instrumentation and Electrical Engineering, Jilin University, Changchun, China \\
$^2$ Institute of Geophysics, University of Tehran, Iran \\
$^3$ School of Mathematical and Statistical Sciences, Arizona State University, Tempe, AZ, 85287, USA \\
$^4$  Department of Geography, Geology and Planning,
Missouri State University, Springfield, Missouri, USA.}
\begin{document}
\maketitle

\begin{summary}
A fast algorithm for the large-scale joint inversion of gravity and magnetic data is developed. The algorithm uses a nonlinear Gramian constraint to impose correlation between the density and susceptibility of the reconstructed models. The global objective function is formulated in the space of the weighted parameters, but the Gramian constraint is implemented in the original space, and the nonlinear constraint is imposed using two separate Lagrange parameters, one for each model domain. This combined approach provides more similarity between the reconstructed models. It is assumed that the measured data are obtained on a uniform grid and that a consistent regular discretization of the volume domain is imposed. Then, the sensitivity matrices exhibit a block Toeplitz Toeplitz block structure for each depth layer of the model domain, and both forward and transpose operations with the matrices  can be implemented efficiently using two dimensional fast Fourier transforms. This makes it feasible to solve for large scale problems with respect to both computational costs and memory demands, and to to solve the nonlinear problem  by applying iterative methods that rely only on matrix vector multiplications. As such, the use of the regularized reweighted conjugate gradient algorithm, in conjunction with the structure of the sensitivity matrices, leads to a fast methodology for large-scale joint inversion of geophysical data sets. Numerical simulations demonstrate that it is possible to apply a nonlinear joint inversion algorithm, with $L_p$-norm stabilisers,  for the reconstruction of large model domains on a standard laptop computer. It is demonstrated, that while the $p=1$ choice provides sparse reconstructed solutions with sharp boundaries, it is also possible to use $p=2$  in order to provide smooth and blurred models. The methodology is used for inverting gravity and magnetic data obtained over an area in northwest of Mesoproterozoic St. Francois Terrane, southeast of Missouri, USA.   
\end{summary}

\section{Introduction}\label{sec:intro}
The joint inversion of multiple data sets has recently received considerable attention by the geophysical community. Different geophysical data sets acquired in a given survey area provide information about different physical properties of subsurface target(s). A model consistent with multiple such data sets is, therefore,  more reliable than a model which is produced by only a single data set \cite{Leli:12}. Several approaches have been developed for the joint inversion of multiple geophysical data. Depending on the linkage between different model parameters, these techniques  are usually categorized as applying either petrophysical or structural approaches \cite{FG:09}. 

With a petrophysical approach it is assumed that there is a direct relationship between the different physical properties of the subsurface target(s). This relationship may be a direct theoretical physical relationship,  or it may be  an empirically-derived relationship based on sampled measurements of the rocks \cite{Leli:12}. Although petrophysical linkages are site specific and highly uncertain, there have been many  successful applications  that use these assertions; see for example Nielsen $\&$ Jacobsen \shortcite{NJ:00}, Afnimar et al. \shortcite{Afnimar002}, Leli{\'e}vre et al. \shortcite{Leli:12}, Sun $\&$ Li \shortcite{SL:17}.

A  structural approach for joint inversion of data does not assume any specific analytic relationship between different physical properties. Rather,  it is assumed that the model topology can be used to enhance  structural similarity between reconstructed models \cite{HO:97,FG:09}. Minimizing the value of the cross-gradient between different model parameters, Gallardo \& Meju \shortcite{GM:03},  ensures that changes in most portions of the  different models  occur at similar locations, and that the reconstructed models are structurally similar  \cite{HG:13}. While, the incorporation of the  cross-gradient constraint in an inversion algorithm  has been widely used in the joint inversion literature \cite{GM:03,GM:04,TL:06,FG:09,Moorkamp:2011,HG:13,Gross:19,VLR:2022},  the resulting approach  is blind to  actual constitutive relationships between parameters, and may therefore be of limited applicabilty  \cite{Leli:12}. 

Another approach that also does not require any \textit{a priori} knowledge of specific  relationship(s) between model parameters uses the Gramian determinant constraint. This is calculated as the determinant of the Gram matrix (or matrices for multiple paired data sets) formed for the model parameters, and/or their different attributes \cite{Zh:15}. When the   determinant of the Gram matrix between two sets of parameter vectors is non-zero, the two sets are linearly independent. In contrast, a zero determinant implies linear dependence. Hence by imposing that the determinant is a minimum (here we call this the Gramian constraint), we are explicitly seeking a solution for the model parameters, or their attributes,  which is   correlated, and which is physically relevant.   This methodology, which has been used extensively for the joint inversion of geophysical data sets, see, for example, Zhdanov et al. \shortcite{ZGW:12}, Lin $\&$ Zhdanov \shortcite{LZ:18}, Jorgensen $\&$ Zhdanov \shortcite{JZ:19}, and Zhdanov et al. \shortcite{ZJC:21}, is used here for the large-scale joint inversion of gravity and magnetic data sets.

The  two major computational obstacles for the inversion of  large-scale potential field data sets are the high storage requirements  and the high computational costs  \cite{LiOl:03}. For example, $18$~GBs of memory are required to store a dense matrix of size $15000\times 150000$, corresponding to $15000$ data points and a volume with  $150000$ unknown model parameters, assuming that each real floating point number uses $8$ bytes.  Hence, for joint inversion of multiple data sets with the same volume discretization for the model parameters, the storage scales with the number of data sets. When the memory is sufficient to store the needed matrices, the computational cost of applying matrix operations is high, not only due to the need for a high number of flops\footnote{Floating point operations} per forward or transpose matrix multiply, but also due to the high demand for accessing a memory hogging matrix entries.  These obstacles limit the sizes of the problems which can be solved using current desktop, or   laptop, computers. 

Multiple approaches   have been adopted to mitigate the computational challenges,   including (i) the use of wavelet and compression techniques  to  significantly reduce the size of the sensitivity matrices \cite{PoZh:02,LiOl:03,Vo:15} and (ii) the use of iterative and/or randomization algorithms that project the problem onto a smaller subspace, \cite{OlMcEl:93,OlLi:94,VRA:2017}  and  \cite{VRA:18,V:2020b}, respectively. By designing a   discretization of the volume domain such that the resulting matrices exhibit a structure that is amenable to reduced storage, and also  to efficient forward and transpose computations, it becomes feasible to apply standard iterative and/or randomization algorithms for the solution of large-scale problems  \cite{R:2020}. Specifically, for a uniform volume discretization, the sensitivity matrices exhibit a block Toeplitz Toeplitz block (\bttb) structure for each depth layer of the model,  \cite{ZhangWong:15,ChenLiu:18}, and   both forward and transpose matrix operations can be implemented using the $2-$D fast Fourier transform (\twoDFFT), as described by Vogel  \shortcite{Vogel:2002} and as explicitly explained for the gravity and magnetic forward models by Hogue et al. \shortcite{HRV:20}. Furthermore, rather than storing the dense sensitivity  matrices of sizes $m\times n$, it is sufficient to store a limited number of vectors of size $2m$ that describe the relevant \twoDFFT~arrays. This approach has been used for the inversion of magnetic potential field data, Renaut et al \shortcite{R:2020}, and extended for the nonlinear inversion of joint data sets using the cross-gradient constraint by Vatankhah et al.  \shortcite{VLR:2022}. 

Although, it has been demonstrated that it is possible to solve the nonlinear joint inversion problem with the cross-gradient constraint, the constraint linearization   leads to the inclusion of a large Jacobian matrix in the inversion algorithm, which  limits the efficiency of the  conjugate gradient (\cg) algorithm that is used to solve the resulting symmetric positive definite linear systems  \cite{VLR:2022}. Here we extend the general approach that was used in Vatankhah et al. \shortcite{VLR:2022}, for sensitivity matrices with \bttb~structure, to use the Gramian constraint in combination with the regularized reweighted conjugate gradient (\rrcg) algorithm, Zhdanov \shortcite{Zh:15}, for the solution of the nonlinear joint inversion of magnetic and gravity potential field data. Within the framework of the \rrcg~algorithm matrix inversions are avoided and  a fast methodology can be  obtained. 

Finally, we note that a general inversion of a single geophysical data set requires  the use of a stabilizing functional, which is selected based on the type of model features that one wishes to recover. Such stabilizing functionals are also incorporated in the joint inversion approach, and are applied to each model independently.  Stabilizers that have been widely used in geophysical applications include those that are   based on $L_p$-norms ($p=0$, $1$, $2$) of the model parameters or their gradients. The $L_2$-norm leads to   solutions which are smooth and have minimum structure \cite{CPC:87,LiOl:96,Pi:97}, and can be expected to only provide  the important and large-scale features of  subsurface target(s). To obtain models that are more consistent with true geologic structures, Farquharson  \shortcite{Far:2008}, it is necessary to apply stabilizers that can generate compact or blocky solutions, requiring the use of $p=0$, or $p=1$, for the model or its gradient, respectively,  \cite{LaKu:83,BaSi:94,FaOl:98,PoZh:99,BoCh:2001,Ajo:2007,Far:2008,VRA:2014,SunLi:2014,VRA:2017,FoOl:19}. The generalized framework for the $L_p$-norm stabilizer  provided in  Vatankhah et al. \shortcite{VRL:20a} is easily extended for incorporation in a joint inversion, as was applied for the cross gradient inversion in Vatankhah et al. \shortcite{VLR:2022}. Moreover, the framework makes it trivial to switch between choices for $p$. Here, we use $p=1$ for the norm of the model parameters, so as to generate  compact and sparse solutions with the aim to reconstruct localized sources with sharp boundaries, but note that the choice $p=2$ generates smooth solutions. Hence, the presented  algorithm with the Gramian constraint can be used to produce sparse, or smooth solutions, of the joint inversion of large-scale  geophysical data-sets.

The paper is organized as follows. The  joint inversion algorithm is presented in \cref{methodology}, including the design of the algorithm for use with the Gramian constraint in \cref{joint}, a brief discussion on how the \twoDFFT~is used in \cref{BTTB}, and a discussion of strategies to find the regularization parameters in \cref{regparameter}. The presented algorithm is validated, and the impact of the use of the required parameters is discussed,  for two synthetic examples in~\cref{synthetic}.  In~\cref{real}, practical data sets are inverted for data obtained over an area in northwest of Mesoproterozoic St. Francois Terrane, southeast of Missouri, USA.  Conclusions are presented in~\cref{conclusion}.

\section{methodology}\label{methodology}
Here we provide our general methodology for the joint inversion of gravity and magnetic potential field data sets. We first provide the general joint formulation in \cref{joint} for a general   nonlinear coupling   constraint function $S$. The specifics for the Gramian constraint ($S$ replaced by $S_G$) are discussed in \cref{sec:gramian}, a description of the objective function with a more general weighting formulation is discussed in \cref{sec:modelweight}, a brief derivation of the algorithm is given in \cref{sec:algderiv} and some specific details for the algorithm are given in \crefrange{alg:bounds}{alg:termination}. Implementation aspects, including  the use of the \twoDFFT~implementation for the sensitivity matrices is provided in \cref{BTTB},  and  approaches for determining  the  scalar Lagrange parameters   is given in \cref{regparameter}.  

\subsection{The general joint inversion algorithm}\label{joint}
We consider the widely-used strategy for $3$-D linear inversion of gravity and magnetic data sets. The subsurface is divided into a set of prisms, in which the prisms are kept fixed during the inversion, and the unknown values of the physical properties of the prisms are the model parameters which should be estimated in the inversion \cite{Blakely}. Here, we assume that (i)  the measurements data are obtained on, or can be easily interpolated to,  a uniform grid \cite{BoCh:2001}, (ii) there is no remanent magnetization, and (iii)  self-demagnetization effects are negligible. With these assumptions, the inversion only considers the induced magnetization  and the magnetic anomaly is linearly related to the susceptibility of the prisms. 

In general, we adopt the  use of the superscript $(i)$ on a vector, or a matrix, to indicate the relevant vector, or matrix, for the gravity or magnetic problem, respectively. The vectors $\bfm^{(1)}=(\rho_{1}, \rho_{2},\dots, \rho_{n}) \in \Rn$ and $\bfm^{(2)}=(\kappa_{1}, \kappa_{2},\dots, \kappa_{n}) \in \Rn$ are used for the density and magnetic susceptibility of the prisms and the  data and model parameters are related via   
\begin{align}\label{d=gm}
\bfdo^{(i)}= A^{(i)} \bfm^{(i)}, \,      i=1,2, 
\end{align} 
where the vectors $\bfdo^{(i)} \in \Rm$, $i=1,2$, are the observed gravity and magnetic potential field data. We assume that there are $m$ measurements, and the aim is the reconstruction of the model volume with $n$ unknowns, so that the sensitivity matrix $A^{(i)}$ is of size $m \times n$. The    $A^{(i)}$  are obtained by discretizing the  kernels, for gravity and magnetic problems, respectively, for which we use the formula developed by Haáz \shortcite{Haaz} for computing the vertical gravitational component, and Rao \& Babu \shortcite{RaBa:91} for the total magnetic anomaly, of a
right rectangular prism.

Our goal  is to find  geologically acceptable models $\bfm^{(i)}$ that predict $\bfdo^{(i)}$,  $i=1$, $2$, at the relevant noise levels, by minimizing  the objective function
\begin{multline}\label{GlobalFunction1}
P^{(\alpha,\lambda)}(\bfm^{(1)},\bfm^{(2)}) = \sum_{i=1}^2 {\| \Wd^{(i)}(A^{(i)} \bfm^{(i)}-\bfdo^{(i)}) \|_2^2} + \sum_{i=1}^2 { \alpha^{(i)} \| W^{(i)}(\bfm^{(i)}-\bfma^{(i)}) \|_2^2} \\
+ \lambda~S (\bfm^{(1)},\bfm^{(2)}),  \,  i=1, 2. 
\end{multline}   
Here (i) $S$ is a nonlinear function that imposes correlation between density and susceptibility models,  (ii) $\alpha^{(i)}$ and $\lambda$ are positive scalars (Lagrange parameters) that determine the relative weights of the different terms, (iii) $\bfma^{(i)}$ are initial \textit{a priori} estimates of the models, where it is also possible to set  $\bfma^{(i)}=\bfzero$, (for the zero vector $\bfzero$ of appropriate length), (iv) $\Wd^{(i)}$ are diagonal data whitening  matrices, with  $(\Wd^{(i)})_{jj} = ( \sigma^{(i)}_j)^{-1}$  where $\sigma_j^{(i)}$ are the standard deviations of the error in the data points $(\bfdo^{(i)})_j$, for each $i$ and $j$, and (v) $W^{(i)}$ are diagonal weighting matrices on the model parameters. These are given as a product of diagonal weighting matrices 
\begin{align}\label{W}
W^{(i)}=\Wdepth^{(i)} \Wh^{(i)} \WL^{(i)} \, \in \,\Rnn, \, i=1, 2, 
\end{align}  
where (i)  $\Wdepth^{(i)}$ are standard  depth weighting matrices, see Li \& Oldenburg \shortcite{LiOl:96},  (ii) hard constraint matrices $\Wh^{(i)}$ can be used to assign known values on the model by setting the corresponding diagonal entries to be large so that the known values are fixed during the iterations, where it is also possible to set  $\Wh^{(i)}=\bfeye$, (for the identity matrix $\bfeye$ of appropriate size), and (iv) 
\begin{align}\label{WL}
\WL^{(i)}=\mathrm{diag}\left(   \frac{1}{((\bfm^{(i)}-\bfma^{(i)})^2+\epsilon^2)^{1/4}}   \right) \, i=1,2,
\end{align} 
arises from approximation of the $L_1$-norm stabilizer with a $L_2$-norm term, see Vatankhah et al. \shortcite{VRA:2017} for details. Noting that $\WL^{(i)}$ depend on $\bfm^{(i)}$, the minimization of \cref{GlobalFunction1}, even without the coupling constraint defined by $S$, is nonlinear. Generally, an iteratively re-weighted algorithm is used for the mnimization, in which matrices  $\WL^{(i)}$ are  estimated using the model parameters obtained at the previous iteration, \cite{LaKu:83,PoZh:99,VRA:2017}, and the focusing parameter $1\gg \epsilon>0$ is  added to  assure that each $\WL^{(i)}$ is invertible. Here we us $\epsilon^2=1e^{-9}$. 

\subsection{The Gramian constraint }\label{sec:gramian}
For the nonlinear term $S$ we use the Gramian determinant constraint given by   
\begin{align}\label{Gramian}
S_G (\bfm^{(1)},\bfm^{(2)})=\left\vert   \begin{array}{cc} (\bfm^{(1)},\bfm^{(1)})&(\bfm^{(1)},\bfm^{(2)}) \\ (\bfm^{(2)},\bfm^{(1)})&(\bfm^{(2)},\bfm^{(2)}) \end{array}  \right\vert 
\end{align}
where operator $(.,.)$ indicates the $L_2$ inner product  \cite{ZGW:12,Zh:15}. When the model parameters are linearly related via $\bfm^{(1)}=\gamma \bfm^{(2)}$, 
for a scalar $\gamma$, $S_G=0$. The minimization of \cref{GlobalFunction1} for $S=S_G$   enforces, therefore,  linear correlation of the model parameters, and no  \textit{a  priori} knowledge about model correlation is required, which is instead enhanced intrinsically during the inverse iterations. 

Although we apply the Gramian constraint for the  model parameters,  it is also possible to use other transformations of the model parameters \cite{Zh:15}. For example, if we use the gradient of the model parameters, then the Gramian constraint enhances the correlation between the gradients. In this case, the gradients are enforced to be  parallel, and the constraint is similar to that which is imposed by using the   cross-gradient function \cite{Zh:15}. As presented, the  Gramian constraint provides a general approach for the joint inversion of multiple geophysical data sets.

\subsection{Reformulation of the objective function}\label{sec:modelweight}
The objective function given in \cref{GlobalFunction1}  can be reformulated using a space of weighted parameters as
\begin{multline}\label{GlobalFunction2}
P^{(\alpha,\tilde{\lambda})}(\mt^{(1)},\mt^{(2)})=\sum_{i=1}^2 {\| \At^{(i)} \mt^{(i)}-\dotilde^{(i)} \|_2^2} +  \sum_{i=1}^2 { \alpha^{(i)} \| \mt^{(i)}-\mat^{(i)} \|_2^2}  + \\ \tilde{\lambda}~\tilde{S}_G (\mt^{(1)},\mt^{(2)}). 
\end{multline}
Here,  we introduce the weighted variables
\begin{align}\label{weighting}
\At^{(i)} = \Wd^{(i)} A^{(i)} (W^{(i)})^{-1},  \,  \dotilde^{(i)}=\Wd^{(i)} \bfdo^{(i)}, \,
\mt^{(i)}=W^{(i)} \bfm^{(i)}, \, \mat^{(i)}= W^{(i)} \bfma,
\end{align}
and explicitly introduce the notation $\tilde{\lambda}$ and $\tilde{S}_G$ to indicate that the coupling constraint and the Lagrange parameter have  a different weighting in the weighted variables, as shown in \cref{steepest} in \cref{GramianA2} we do not have directly $\tilde{S}=W S$ for a diagonal weighting matrix $W$. 
Although \cref{GlobalFunction2} shows the Gramian constraint in the space of weighted parameters, in our implementation it is imposed in the original (unweighted) space so as to  provide the correlation between the original model parameters and not their weighted forms.  We demonstrate in~\cref{steepest}, that this does not require any significant change in the implementation of the algorithm.

\subsection{The regularized re-weighted conjugate gradient algorithm}\label{sec:algderiv}
We now provide the essential components for minimizing \cref{GlobalFunction2}, assuming given initial values for all variables.   Additional details  are presented,  as needed, in \cref{steepest}. 
The first variation of the parametric functional \cref{GlobalFunction2}  is given by, \cite{Zh:15},
\begin{multline}\label{FirstVariation}
\delta P^{(\alpha,\tilde{\lambda})}(\mt^{(1)},\mt^{(2)}) = 2 \sum_{i=1}^2 (\At^{(i)} \delta \mt^{(i)})^{T} (\At^{(i)} \mt^{(i)}-\dotilde^{(i)}) + \\ 2 \sum_{i=1}^2 { \alpha^{(i)} ( \delta \mt^{(i)})^{T} (\mt^{(i)}-\mat^{(i)})}  
+ 2   \sum_{i=1}^2 { \tilde{\lambda}^{(i)} ( \delta \mt^{(i)})^{T}   \tilde{\bfl}_G^{(i)}  }, 
\end{multline}
where $\tilde{\bfl}_G ^{(i)}$ are the   steepest ascent directions of the Gramian constraint $\tilde{S}$   given by 
\begin{subequations}
\begin{align}
\tilde{\bfl}_G^{(1)}&= \|\mt^{(2)} \|_2^2 ~\mt^{(1)} - (\mt^{(1)},\mt^{(2)})~\mt^{(2)}\label{steepestascenta} \\
\tilde{\bfl}_G^{(2)}&= \|\mt^{(1)} \|_2^2 ~\mt^{(2)} - (\mt^{(1)},\mt^{(2)})~\mt^{(1)},\label{steepestascentb} 
\end{align}
\end{subequations}
as shown in~\cref{descentrela}.

\Cref{GlobalFunction2} is minimized using an iterative algorithm. We use a subscript on any variable to indicate the value of that variable at iteration $k$, with $k$ increasing from $0$. Moreover, we introduce the notation that the variable with subscript $k$ but no superscript, and the overbar, corresponds to the block concatenation of appropriate individual variables. For example,  $\bar{\At}_k$  is  the block diagonal obtained from $\At_k^{(1)}$ and $\At_k^{(2)}$, while $\bar{\mt}_k$ is the vector concatenation of the individual vectors $\mt_k^{(1)}$, and $\mt_k^{(2)}$ that is consistent with the definition of $\bar{\At}_k$, and $\bar{\bfl}_k$ is the vector concatenation of the individual vectors $\bfl_k^{(1)}$ and $\bfl_k^{(2)}$. The $k^{th}$ step of the regularized re-weighted conjugate gradient (\rrcg) algorithm  is given by 

\begin{subequations}\label{RRCG}
\begin{align}
\bar{\rtilde}_k &= \bar{\At}_k~\bar{\mt}_k - \bar{\tilde{\bfd}}_\mathrm{obs},   \label{RRCG:1} \\
(\bfl_G^{(1)})_k &= \|\bfm_k^{(2)} \|_2^2 ~\bfm_k^{(1)} - (\bfm_k^{(1)},\bfm_k^{(2)})~\bfm_k^{(2)}, \label{RRCG:2} \\
(\bfl_G^{(2)})_k&= \|\bfm_k^{(1)} \|_2^2 ~\bfm_k^{(2)} - (\bfm_k^{(1)},\bfm_k^{(2)})~\bfm_k^{(1)},  \label{RRCG:3} \\
\bar{\bfl}_k &= \bar{\At}_k^\top~\bar{\rtilde}_k + \bar{\bfalpha}_k ~(\bar{\mt}_k-\bar{\mt}_{k-1}) + \bar{\lambda} ~(\bar{\bfl}_G)_k,   \label{RRCG:4}\\
\bar{\bfp}_k  &= \bar{\bfl}_k + \frac{\| \bar{\bfl}_k \|_2^2}{\| \bar{\bfl}_{k-1} \|_2^2}~\bar{\bfp}_{k-1}, \, \bar{\bfp}_0=\bar{\bfl}_0,   \label{RRCG:5}\\
 {s}_k &= \left( \bar{\bfp}_k,\bar{\bfl}_k \right) /  \left[  \| \bar{\At}_k~\bar{\bfp}_k\|_2^2 ~+ \bar{\bfalpha}_k~\| \bar{\bfp}_k \|_2^2  ~+ \bar{\lambda}~\| \bar{\bfp}_k \|_2^2  \right], \label{RRCG:6}\\
\bar{\mt}_{k+1} &= \bar{\mt}_k -  {s}_k ~ \bar{\bfp}_k \label{RRCG:7}\\
\bar{\bfm}_{k+1}^{(i)}&=(W_k^{(i)})^{-1}~\bar{\mt}_{k+1}^{(i)}. \label{RRCG:8}
\end{align}
\end{subequations}  
As given, the  steepest ascent directions of the Gramian constraint are estimated using the original model parameters, which are easily obtained from the weighted parameters.   

\subsection{Bound Constraints}\label{alg:bounds}
It is standard when handling the ambiguity of the geophysical potential field inversion problem to apply lower and upper bounds on the variables in order to recover a feasible image of the subsurface. Here the constraints are imposed at each iteration of the inversion
process.  If an estimated density or magnetic susceptibility value of a given prism falls outside the specified bounds, the value for that prism is projected to the nearest bound \cite{LaKu:83,PoZh:99,BoCh:2001}. 
\subsection{Termination of the Algorithm}\label{alg:termination}
To terminate the inversion we apply the standard $\chi^2$-test that the models predict the data  to the given noise level, specifically that 
\begin{align}\label{chi2}
\Phi(\bfd^{(i)}_{k+1}) \leq m+\sqrt{2m}, \end{align} 
 for $i=1$ and $i=2$, where 
\begin{align*}
\Phi(\bfd^{(i)}_{k+1})=\| \Wd^{(i)}(A^{(i)} \bfm_{k+1}^{(i)}-\bfdo^{(i)}) \|_2^2.
\end{align*}
An upper limit, $K_{\mathrm{max}}$, on the number of iterations to be performed is also imposed, in case the $\chi^2$ value is not achieved after a suitable number of steps. This may occur if the arbitrary scalar parameters need to be adjusted, as discussed in \cref{regparameter}. 

\subsection{The BTTB structure of the sensitivity matrices} \label{BTTB}
From the steps of the \rrcg~algorithm given in \crefrange{RRCG:1}{RRCG:8}, it is immediate that the iteration proceeds using only forward and transpose matrix vector multiplications with the respective matrices, and no matrix inversion is required. Hence, all such operations can proceed by using the \bttb~structure of these matrices, as described in detail in Hogue et al. \shortcite{HRV:20}, where downloadable software is also given for implementing these operations. This was also used for the joint inversion of gravity and magnetic data, in conjunction with the cross-gradient constraint, by Vatankhah et al \shortcite{VLR:2022}. Hence the \rrcg~ algorithm provides a fast and minimal memory  strategy for minimizing the global objective function~\cref{GlobalFunction2} for large-scale problems. We give a brief overview of the use of the \bttb~matrices in~\cref{BTTBdetail}, but first note that the requirement to use a regular grid with constant cell sizes is in no way limiting, since data that are provided away from a regular grid can be interpolated to the appropriate points, as needed. Moreover,  the regular grid assumption is standard, Boulanger \& Chouteau \shortcite{BoCh:2001}, and the design of the volume discretization is independent of how the data are provided; it is standard to assume a uniform volume discretization.

\subsubsection{Matrix operations using the \twoDFFT}\label{BTTBdetail}
Suppose that the measurement stations are located at the center of the prisms on the surface of the domain which is potentially padded in East and North directions by extra prisms in which there are no measurements \cite{BoCh:2001}. For the general case, independent of gravity or magnetic data, we assume that  the entries of the sensitivity matrix $A$ correspond to the contribution from prisms with respect to measurement stations. This matrix is  decomposed into column blocks,  $A=\left[A_{(1)}, \dots, A_{(n_z)} \right]$, where block $A_{(r)}$ corresponds to the $r^{th}$ depth level for a total of $n_z$ blocks, as given by 
for the $n_z$ depth layers in the model. Regardless of the structure of each block, the  multiplication of matrix $A$ with a vector $\bfx \in \Rn$ uses the column block structure of $A$;  $A \bfx =\sum_{r=1}^{n_z} {A_{(r)}} \bfx_{(r)}$ where $\bfx$ is blocked consistently with $A$. Thus, it is only necessary to determine the efficient evaluation of a single term in the sum. 

The blocks of the sensitivity matrices for the gravity and magnetic problems each exhibit block-Toeplitz-Toeplitz-block (\bttb) structure;  symmetric for the gravity problem and unsymmetric for the magnetic problem.  In either case, the block \bttb~matrix can be embedded into a block Circulant Circulant block (\bccb) matrix which facilitates the use of the \twoDFFT~ for all matrix-vector operations   \cite{Vogel:2002}. To perform the embedding one needs only the first column, or first row, or for the unsymmetric case both the first row and the first column, in order to fully describe the underlying operation with the matrix as a \twoDFFT. 

Consider the case where we use just the first column, then this is extended and reshaped into an array $T \in \Rsxy$. Here $(s_x,n_x)$ and $(s_y,n_y)$ indicate the number of stations and prisms in $x$ and $y$ directions, respectively. The dimension of array $T$ is much smaller than the corresponding  block $A_{(r)}$, and thus the storage of $A$ is replaced by that of $n_z$ much smaller matrices,  $T_{(r)}$.   Now to find $A_{(r)} \bfx_{(r)}$, $\bfx_{(r)}$ is reshaped and embedded into the upper left corner of a zero matrix $X \in \Rsxy$ \cite{HRV:20}. Then $A_{(r)} \bfx_{(r)}$ is extracted from
\begin{align}
\iffttwo(\ffttwo(T)\odot \ffttwo(X)),
\end{align} 
where $\odot$ represents element-wise multiplication, and \ffttwo~ and \iffttwo~ indicate the two-dimensional FFT and inverse two-dimensional FFT, respectively. It is immediate that a set of equivalent steps can be used to calculate $(A_{(r)})^\top \bfv$, since $(A_{(r)})^\top$ is also \bttb. Therefore, multiplication using the transpose matrix uses
 \begin{align}
\iffttwo(\texttt{conj}(\ffttwo(T)).*\ffttwo(V)).
\end{align} 
Furthermore, the \rrcg~ algorithm requires also matrix products with $\At$ or $\At^\top$. But the weighting matrices in $\At=\Wd  A  W^{-1}$ are diagonal, and thus  
\begin{align}\label{matrixvector}
\At \bfx =\Wd \odot ( A  (W^{-1} \odot \bfx)),
\end{align}
where we consider the diagonal weighting matrices as vectors of the appropriate lengths. The
multiplication with the diagonal matrices is insignificant and the \twoDFFT~ is only needed to evaluate $\At \bfw$ where $\bfw = W^{-1}\odot \bfx$. The same strategy is used for the transpose multiplication \cite{HRV:20}.

\subsection{Estimating regularization/Lagrange parameters}\label{regparameter}
Generally, the parameters $\alpha^{(i)}$ and $\lambda^{(i)}$ that determine the relative weights for the stabilizers and the Gramian constraint, are referred to as regularization and Lagrange parameters, respectively. The choice of these parameters is significant for the algorithm and impacts the convergence rate of  the joint inversion algorithm.
For ill-conditioned systems, as here for the potential field problems given by \cref{d=gm}, it is well known that  small values of $\alpha^{(i)}$ lead to unstable solutions, but large values will imply that the data misfit terms are not significant to the minimization and poor data fits of the observed and predicted data  will be obtained. On the other hand, if the $\lambda^{(i)}$ are small, the constraints are only weakly applied, and there will be limited correlation between the the two sets of model parameters.  It is clear, that all parameters have to be chosen carefully. 

For independent inversion of individual geophysical data, it is possible, and feasible,  to apply automatic parameter-choice methods to estimate an optimum value for the corresponding single regularization parameter, see for example Vatankhah et al. \shortcite{VRA:2014}. Here, for the joint inversion, four parameters are needed, $\alpha^{(i)}$ and $\lambda^{(i)}$ for $i=1$, $2$, and  is difficult, or potentially impossible to design an automatic and efficient procedure for finding optimal choices. Here, we use an effective and simple strategy to determine these parameters. At the first iteration,  large values for $\alpha^{(1)}$ and $\alpha^{(2)}$ are selected. Then, the  parameters are reduced slowly according to 
\begin{align}\label{regpara}
\alpha_{k+1}^{(i)}=\alpha_{k}^{(i)} q^{(i)}, \, i=1,2. 
\end{align}  
where it is explicit that the decay rate parameter $q^{(i)}$ can be different for each problem.    The process continues until the predicted data of one of the reconstructed models satisfies the observed data at the noise level as given in \cref{chi2}. For that data set, the relevant parameter is then kept fixed during the subsequent iterations.

Lagrange parameters, $\lambda^{(1)}$ and  $\lambda^{(2)}$,  are equally important to the success of the algorithm in obtaining useful models. Here, we will use fixed values for the Lagrange parameters, as chosen for each specific problem and discussed in \cref{synthetic}. This approach means that we can gradually increase the impact of the Gramian constraint, as the $\alpha_k^{(i)}$ decrease with $k$. The results presented in \cref{synthetic} demonstrate that the strategy works well and can provide convergence for the iterative joint inversion algorithm.

\section{Validation of the Algorithm for Synthetic Data}\label{synthetic}
Before implementing the algorithm, as developed in \cref{methodology}, for the inversion of real data sets, it is important to validate the approach on synthetic models with known configurations. We  first evaluate the  performance of the algorithm described by \crefrange{RRCG:1}{RRCG:8}  for the reconstruction of a synthetic model of moderate size consisting of two cubes, as detailed in \cref{cube}. Because the problem is relatively small, it is feasible and efficient  to evaluate the algorithm under different configurations, in particular for different choices of the Lagrange parameters $\lambda^{(i)}$, \crefrange{nogramian}{simLagrange}. Moreover, it is easy to change between using $S_G$ and $\tilde{S}_G$  in the implementation, since it is just required to calculate the search direction calculations using \cref{RRCG:2,RRCG:3} to using \cref{steepestascenta,steepestascentb}. Results illustrating the impact of using this switch are given in \cref{Gramianweightedspace}.
 We also demonstrate the results which are obtained using the joint smoothing algorithm, in which $\WL$ in~\cref{W} is replaced by the identity matrix, yielding the algorithm that implements the Gramian constraint but for a solution with a minimum norm, \cref{grammian:smoothing}. The ease with which this is possible makes it straightforward to compare the more general focusing inversion algorithm, implemented for the $L_1$-norm, with the smoothed solution. 
As a second example we validate the algorithm with focusing inversion and the Gramian constraint, for the solution of a large-scale problem for a model with multiple bodies and a more complicated structure, as described in~\cref{multiple}. 

For both the smaller and multiple models, in the simulation of the total field anomalies, the intensity of the geomagnetic field, the inclination, and the declination are selected as $50000$ nT, $45^{\circ}$, and $45^{\circ}$, respectively. Further, Gaussian noise with zero mean and $j^{th}$ standard deviation,   
\begin{align}\label{noise}
\sigma_j^{(i)}=\tau_1 ~ | (\bfde^{(i)})_j |+\tau_2 ~ \mathrm{max} | \bfde^{(i)} |, \quad j=1 \dots m,
\end{align} 
is  added to the exact   measurements, $(\bfde^{(i)})_j$, with the pairs $(\tau_1, \tau_2)$  selected as $(0.02, 0.01)$ and $(0.01, 0.01)$ for the gravity and magnetic problems, respectively. For the simulated models, the density contrast of each structure is $1$~g~cm$^{-3}$,  the magnetic susceptibility is $0.1$ in SI units, and it is assumed that the structures are embedded in a  homogeneous non-susceptible background. 

In the algorithm,  the initial density and susceptibility models,  $\bfma^{(i)}$, $i=1$, $2$, are set to $\bfzero$, and, except for the investigation of the smoothing norm in~\cref{grammian:smoothing},   the initial parameters $\alpha_{1}^{(1)}$ and $\alpha_{1}^{(2)}$ are set to  $20000$, with the decay rate parameter in \cref{regpara} set to $q=0.95$ for both cases, and the iteration is terminated for $K_{\mathrm{max}}=150$ if the $\chi^2$ test as defined by~\cref{chi2} is not achieved. In all cases, upper and lower bounds, $0\leq \bfm^{(1)}\leq 1$~g~cm$^{-3}$ and $ 0 \leq \bfm^{(2)} \leq 0.1$,  in   SI units,  are imposed at all iterations of the algorithm, and we set $\Wh^{(i)}=\bfeye$, $i=1$, $2$ in~\cref{W}.   To evaluate the performance of the algorithm for the specific parameter settings and configurations  we compare the relative errors of the reconstructed models, as defined by 
\begin{equation}\label{RE}
RE^{(i)}=\frac{\|\bfme^{(i)}-\bfm_{K}^{(i)}  \|_2}{\|\bfme^{(i)} \|_2}, \, i=1, \, 2.
\end{equation}
Here $\bfme$ is the exact model and $\bfm_{K}$ is the reconstructed model at the final iteration $K\le K_{\mathrm{max}}$. The use of the relative error provides a measure that is comparable for assessing the accuracy of both the density and magnetic susceptibility models. All computations are performed on a standard laptop computer with Intel(R) Core(TM) i7-10750H CPU  $2.6$~GHz processor and $16$~GB RAM. The results presented in \crefrange{cube}{multiple} demonstrate that the algorithm is efficient for the solution of large-scale problems  and that acceptable solutions that are correlated are provided. 

\subsection{Two cubes}\label{cube}
We consider a model that consists of two cubes of the same size, where the lengths in East, North and depth directions are $1000$~m, $500$~m and $300$~m, respectively, but placed at different depths, with the top of each cube at depth $100$~m (for the left cube)  and $200$~m (for the right cube), as illustrated in \cref{fig1}. 
The gravity and magnetic data sets are  generated on the surface over a $50 \times 30$  uniform grid with $100$~m grid spacing. The resulting noise-contaminated data sets, using the noise model described in~\cref{noise}, are illustrated in~\cref{fig2}.
To perform the inversion, the model region of depth $1000$~m is discretized into $ 50 \times 30 \times 10 = 15000$ prisms of size $100$~m in each dimension.  

\begin{figure*}
\subfigure{\label{1a}\includegraphics[width=.7\textwidth]{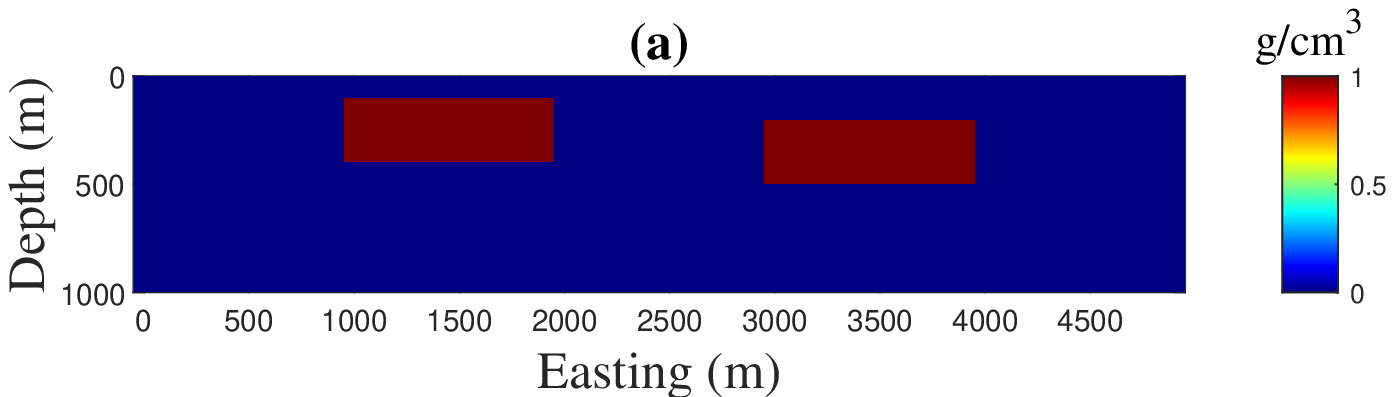}}
\subfigure{\label{1b}\includegraphics[width=.7\textwidth]{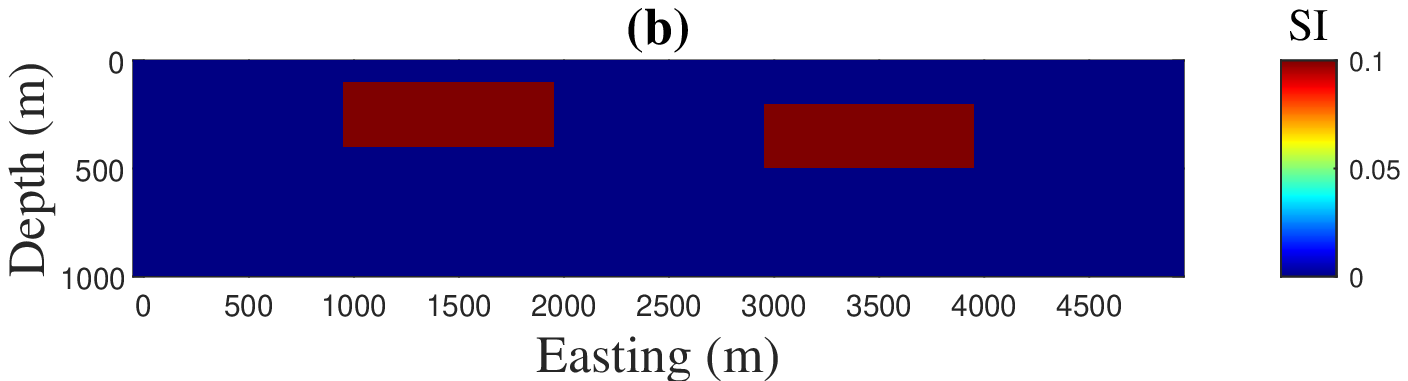}}
\caption {Cross sections of the first synthetic model consisting of two similar cubes located at different depths. In~\cref{1a} the density distribution,  and in  \cref{1b} the magnetic susceptibility distribution.} \label{fig1}
\end{figure*}

\begin{figure*}
\subfigure{\label{2a}\includegraphics[width=.45\textwidth]{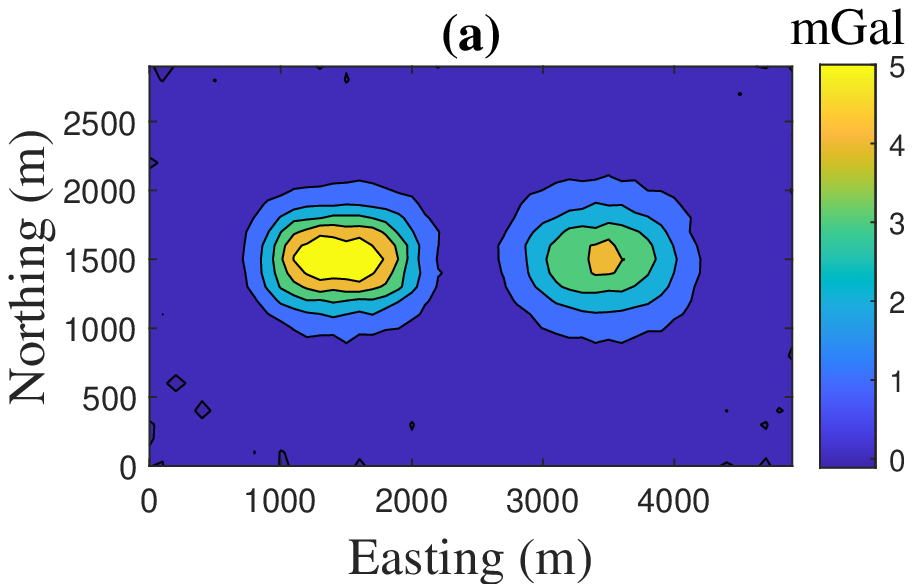}}
\subfigure{\label{2b}\includegraphics[width=.47\textwidth]{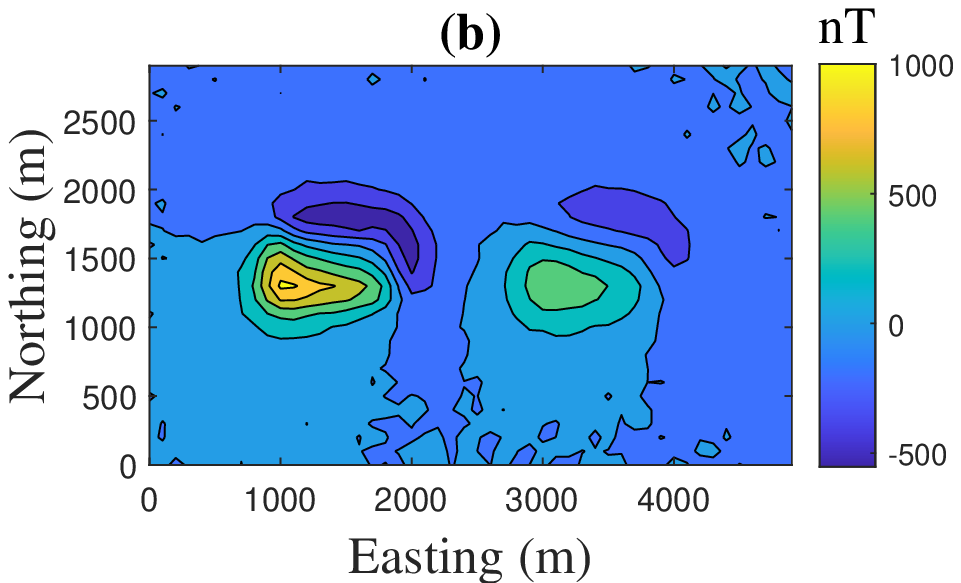}}
\caption{Noise-contaminated data produced by the models shown in~\cref{1a,1b}. In~\cref{2a} the vertical component of the gravity field,  and in \cref{2b} the total magnetic field.} \label{fig2}
\end{figure*}

Three different choices for the Lagrange parameters, $\lambda^{(1)}$ and $\lambda^{(2)}$, are considered for testing the inversion. In each case we assume $\lambda^{(1)} \leq \lambda^{(2)}$, which is needed to provide an equivalent constraint for each model parameter set,  noting that the amplitude of the  magnetic susceptibility is less than the density. All other parameters of the inversion are the same for these three tests, \crefrange{nogramian}{diffLagrange}

\subsubsection{Inversion without the Gramian constraint: $\lambda^{(1)}= \lambda^{(2)}=0$}\label{nogramian}
We first test the performance of the algorithm without the Gramian constraint coupling which is achieved by taking the Lagrange parameters to be zero in each case. When this is implemented within the \rrcg~algorithm, this corresponds to taking the directions in~\cref{RRCG:2,RRCG:3} to be zero. It does not, however, immediately correspond to  the inversion of each data set independently. Specifically the search directions and search parameter in~\cref{RRCG:5,RRCG:6}, respectively, are coupled, unless the search directions are defined independently with the scaling calculated for each model parameter. Here, we explicitly treat the \rrcg~algorithm as a \textit{stand-alone} algorithm for each parameter set.

After $50$ iterations, and only about $43$~s, the convergence criteria are satisfied and the inversion terminates.  The reconstructed density and magnetic susceptibility models, which are geophysically acceptable,  are illustrated in~\cref{3a,3b}.  The models are, however, not similar. The reconstructed cubes of the susceptibility model are aligned with the original upper depths of  the assumed true cubes, but each cube is thicker. In contrast, the  thickness of the  reconstructed density cubes is approximately consistent with the assumed true cubes, but the right cube is not correctly aligned at the upper depth with the true model. The relative errors of the reconstructed density and susceptibility models are $RE^{(1)}=0.5686$ and $RE^{(2)}=0.8198$, respectively, and this quantitative difference is indicative of the results illustrated in~\cref{fig3}. To further examine the quality of the results without the imposition of the coupling constraint $S_G$,  \cref{fig4} illustrates the cross-correlation of the reconstructed density and magnetic susceptibility models. From this plot, it is clear that the reconstructed model parameters are not correlated, which is indicative that the coupling constraint has not been imposed. Finally, \cref{fig5} illustrates the predicted gravity and magnetic data sets which are in  good agreement with the observed data at the noise level.

\begin{figure*}
\subfigure{\label{3a}\includegraphics[width=.7\textwidth]{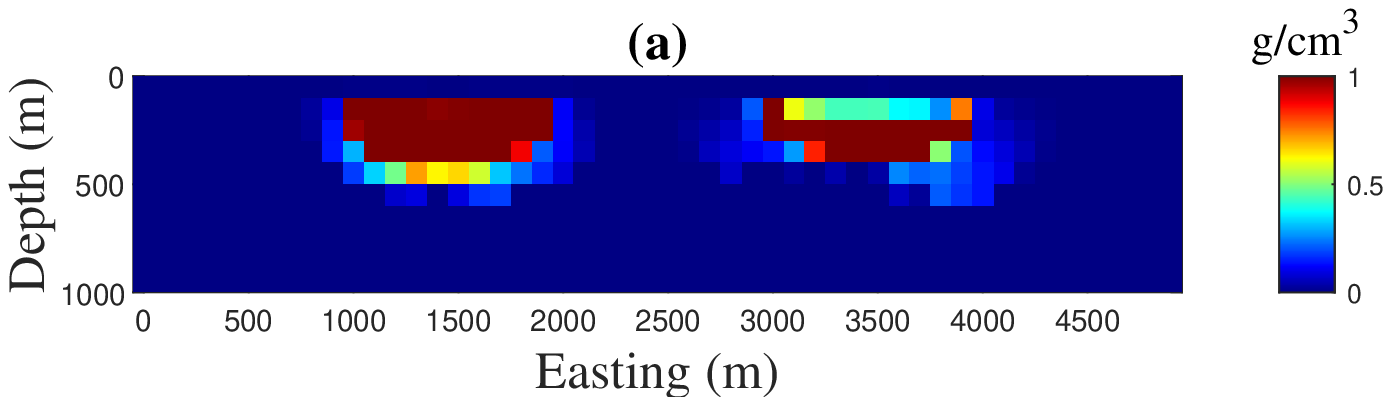}}
\subfigure{\label{3b}\includegraphics[width=.7\textwidth]{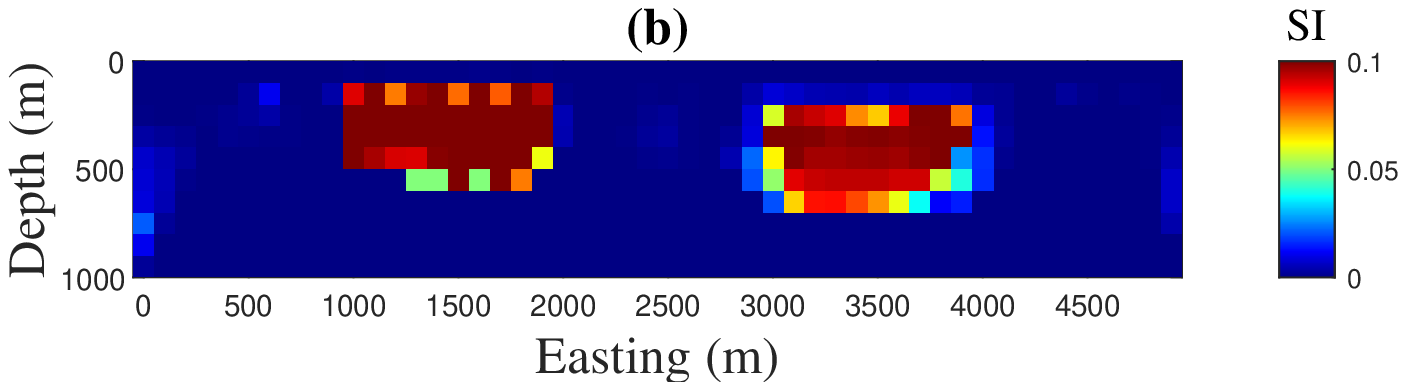}}
\caption {Cross sections of the reconstructed models for the data sets shown in~\cref{2a,2b} for Lagrange parameters set to $0$,   $\lambda^{(1)}=\lambda^{(2)}=0$. In \cref{3a} the  density distribution, and in \cref{3b} the magnetic susceptibility distribution.} \label{fig3}
\end{figure*}

\begin{figure*}
\includegraphics[width=0.5\textwidth]{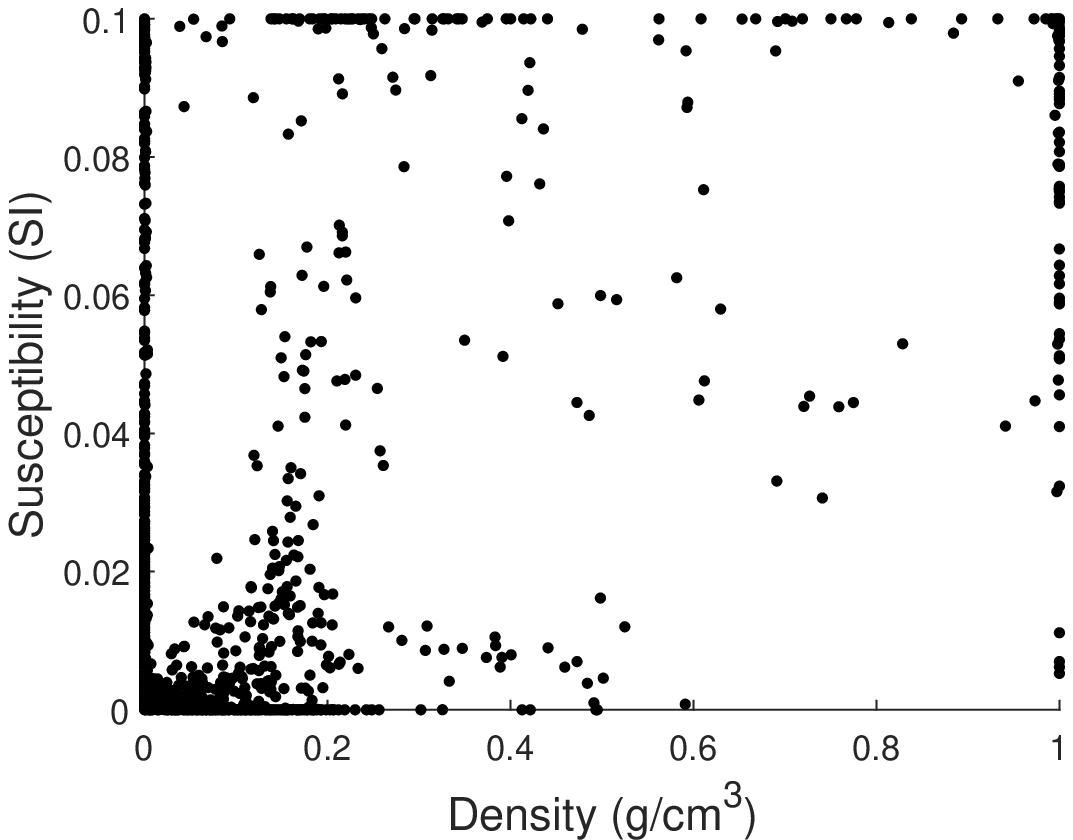}
\caption{The cross-correlation plot between the density and magnetic susceptibility model parameters for the solutions presented in~\cref{3a,3b}, obtained with $\lambda^{(1)}= \lambda^{(2)}=0$.} \label{fig4}
\end{figure*}

\begin{figure*}
\subfigure{\label{5a}\includegraphics[width=.45\textwidth]{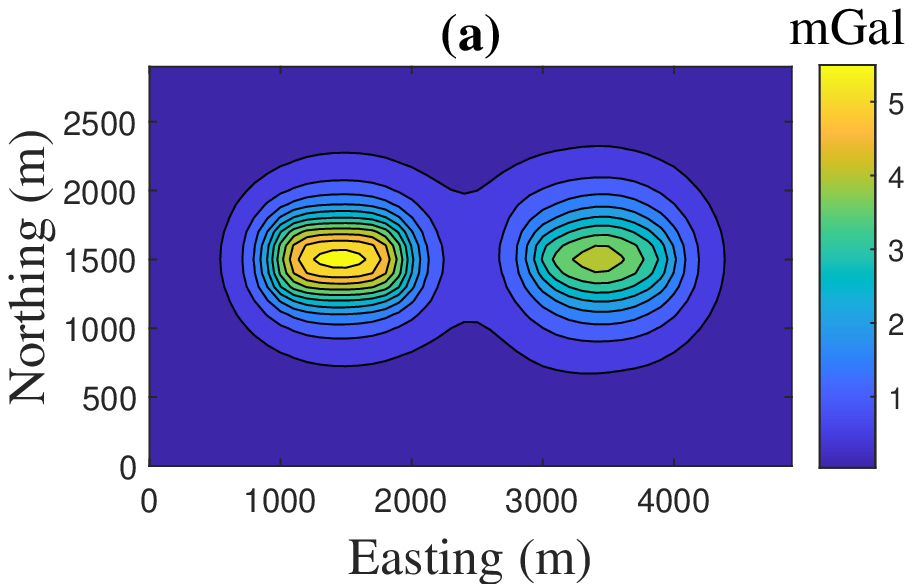}}
\subfigure{\label{5b}\includegraphics[width=.47\textwidth]{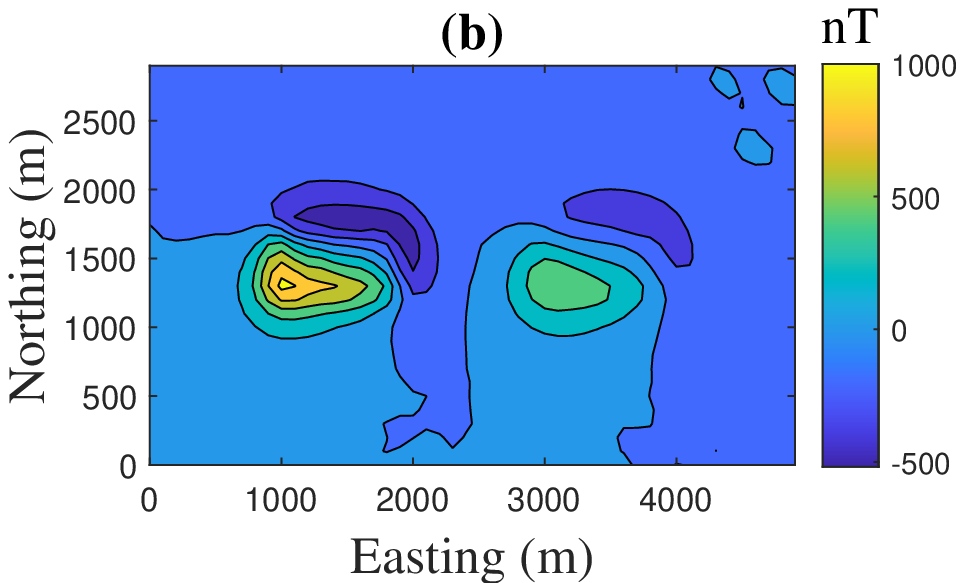}}
\caption{Data predicted by the models shown in~\cref{3a,3b}, obtained with $\lambda^{(1)}= \lambda^{(2)}=0$. In \cref{5a} the  vertical component of the gravity, and in \cref{5b} the total magnetic field.} \label{fig5}
\end{figure*}

\subsubsection{Inversion with Gramian constraint: $\lambda^{(1)}=50$,  and $ \lambda^{(2)}=1500$}\label{diffLagrange}
Here we use Lagrange parameters of moderate size, $\lambda^{(1)}=50$, $\lambda^{(2)}=1500$.  After $52$ iterations, and only about $45$~s, the convergence criteria are satisfied and the inversion terminates, indicating that the Gramian constraint increases the number of iterations slightly, and the computational time accordingly.  The reconstructed density and magnetic susceptibility models, which are  illustrated in~\cref{6a,6b},  are now not only geophysically acceptable but are also  similar and consistent with the true model. Furthermore, the relative errors are comparable,   $RE^{(1)}=0.4761$ and $RE^{(2)}=0.5192$ and the correlation between the model parameters is increased, as illustrated in the cross-correlation plot in~\cref{fig7}. As for the uncoupled case,  the predicted gravity and magnetic data sets, illustrated in~\cref{fig8},  are in  good agreement with the observed data at the noise level.

\begin{figure*}
\subfigure{\label{6a}\includegraphics[width=.7\textwidth]{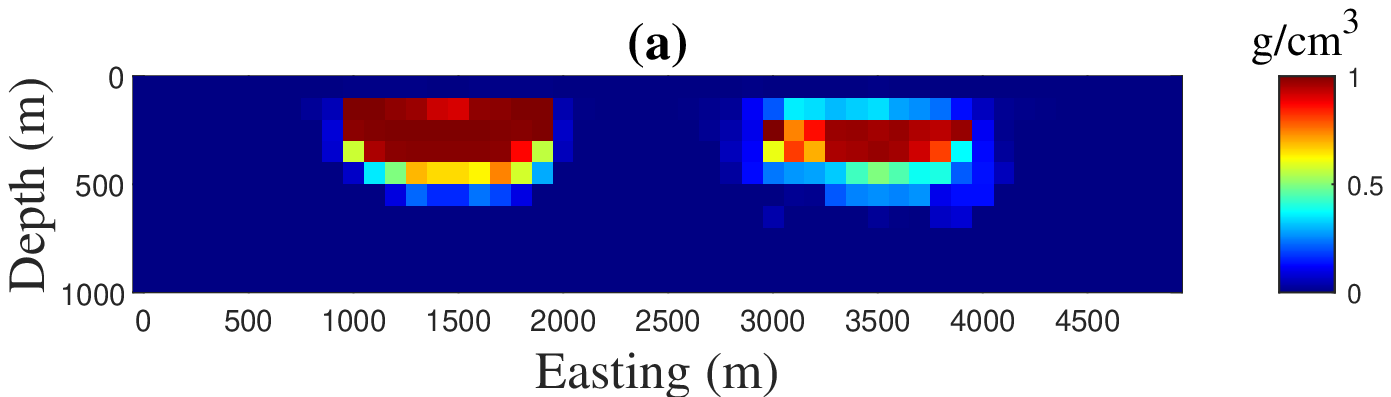}}
\subfigure{\label{6b}\includegraphics[width=.7\textwidth]{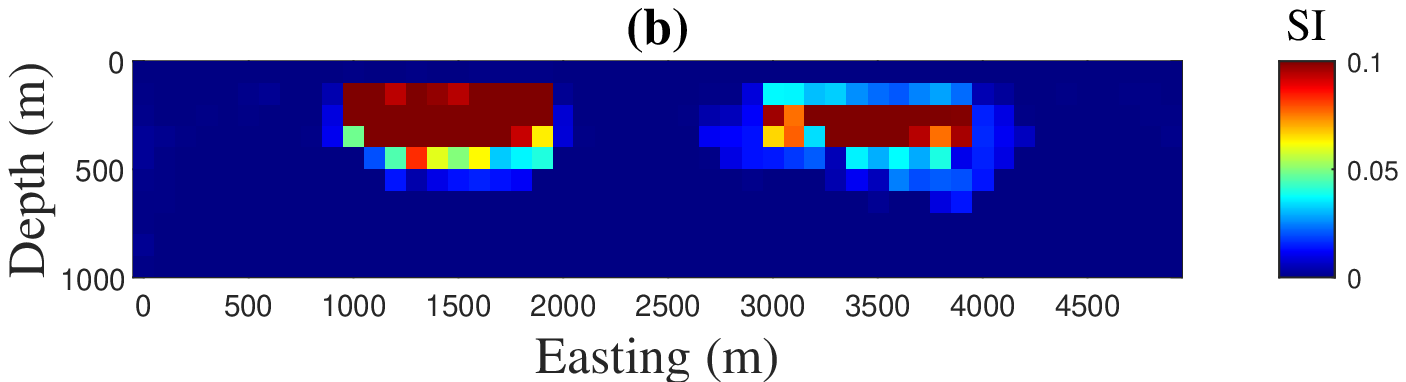}}
\caption {Cross sections of the reconstructed models for the data sets shown in~\cref{2a,2b} obtained with  $\lambda^{(1)}=50$ and $ \lambda^{(2)}=1500$. In \cref{6a} the  density distribution, and in \cref{6b} the magnetic susceptibility distribution.} \label{fig6}
\end{figure*}

\begin{figure*}
\includegraphics[width=0.5\textwidth]{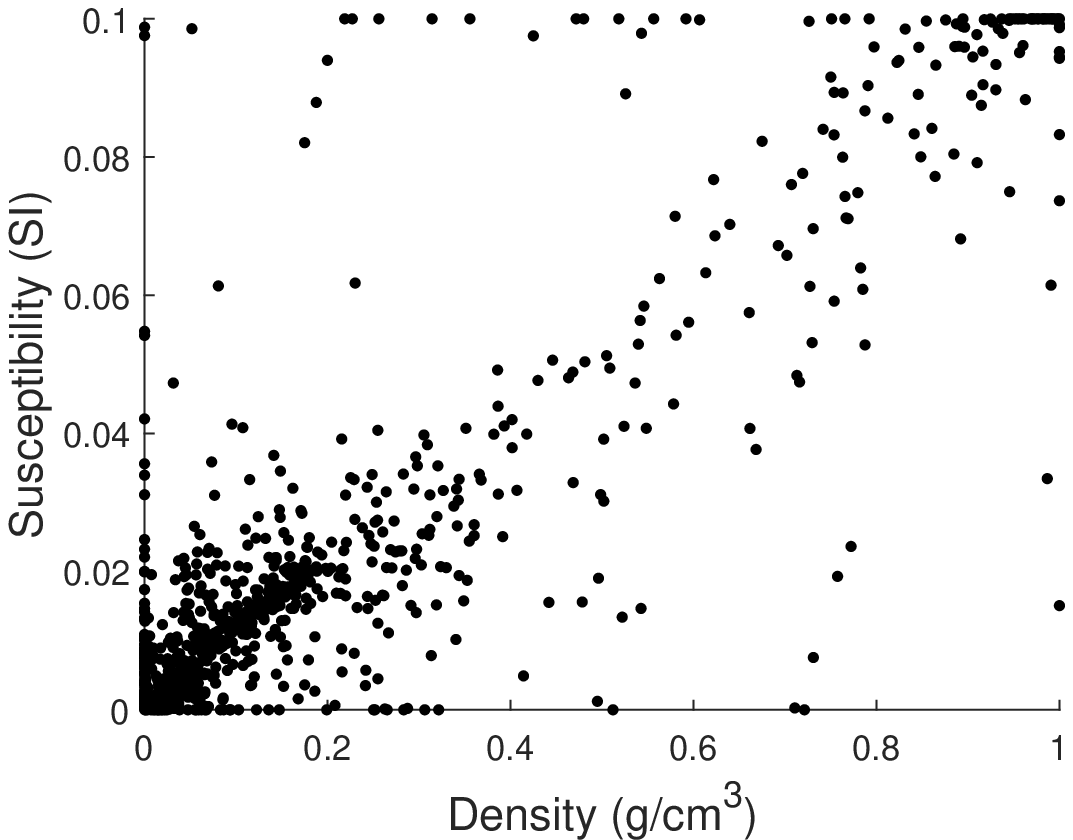}
\caption{The cross-correlation plot between the density and magnetic susceptibility model parameters for the solutions presented in~\cref{6a,6b}, obtained with $\lambda^{(1)}=50$ and $ \lambda^{(2)}=1500$.} \label{fig7}
\end{figure*}

\begin{figure*}
\subfigure{\label{8a}\includegraphics[width=.45\textwidth]{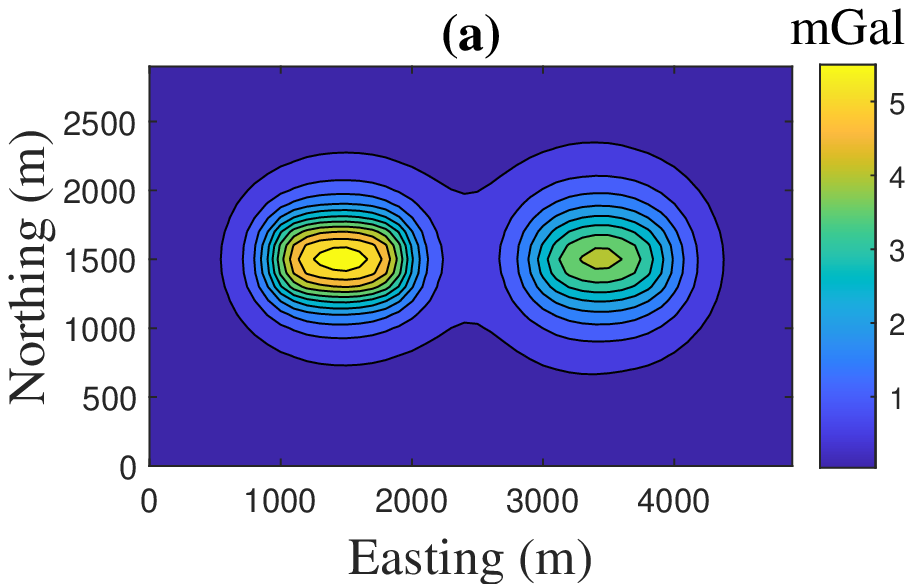}}
\subfigure{\label{8b}\includegraphics[width=.47\textwidth]{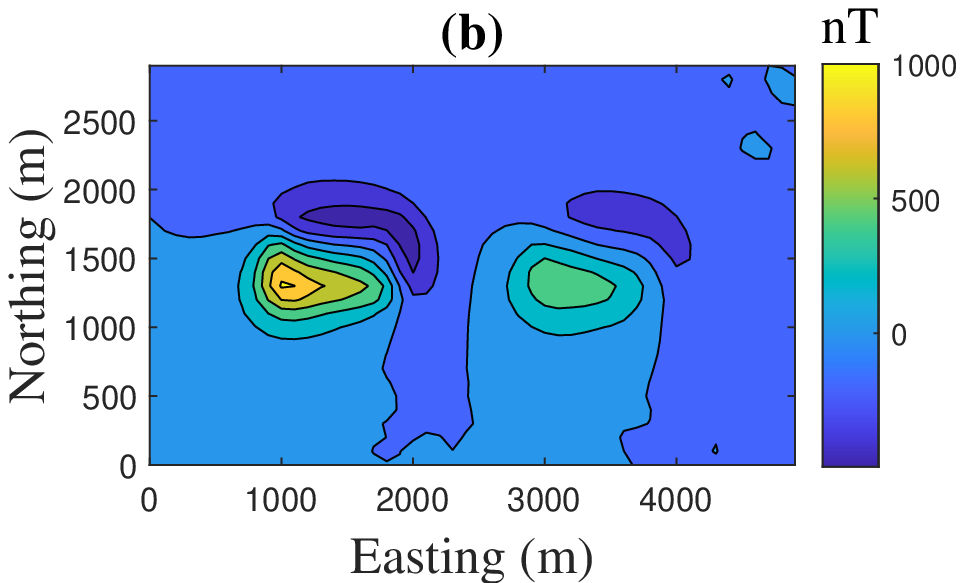}}
\caption {Data predicted by the models shown in~\cref{6a,6b}, obtained with  $\lambda^{(1)}=50$ and $ \lambda^{(2)}=1500$. In \cref{8a} the  vertical component of the gravity, and in \cref{8b} the total magnetic field.} \label{fig8}
\end{figure*}

\subsubsection{Inversion with Gramian constraint: $\lambda^{(1)}=1000$,  and $ \lambda^{(2)}=15000$}\label{simLagrange}
Here we use much larger Lagrange parameters, $\lambda^{(1)}=1000$, $\lambda^{(2)}=15000$.  The convergence criteria are not satisfied,  the iteration terminates at the upper limit of the number of iterations, $K_{\mathrm{max}}=150$, and the computational time is increased accordingly to about $134$~s.
The reconstructed density and magnetic susceptibility models, which are  illustrated in~\cref{9a,9b},  are now  similar but they are not consistent with the true models.  The relative errors,  $RE^{(1)}=0.5715$ and $RE^{(2)}=0.5860$, are comparable, and the cross-correlation plot in~\cref{fig10} suggests that the obtained models are reasonably correlated. Moreover, the  predicted gravity and magnetic data sets illustrated in~\cref{fig11}  are again in  good agreement with the observed data at the noise level. On the other hand, the obtained right cube is not sufficiently focused. 

The presented results are consistent with our expectations for the implementation of the coupling constraint; when the Lagrange parameters are chosen \textit{too} large the constraint is strongly imposed and the solutions obtained are correlated, but the focusing norm has less impact on the solutions, and while the results suggest a reasonable relative error, the lack of convergence is indicative that the algorithm has converged to a solution that is not suitable. This conclusion suggests the process for determining suitable values for the Lagrange parameters. If they are too small the solutions are not well-correlated but when they are too large the coupling dominates and the solutions obtained are not satisfactory. Hence this suggests that the algorithm should be implemented with fixed parameters, but increasing for a number of different runs of the algorithm. The smallest set of $\lambda^{(i)}$ which provide similarity, as indicated by a cross correlation plot, and provide good predictions for the data should be selected.

\begin{figure*}
\subfigure{\label{9a}\includegraphics[width=.7\textwidth]{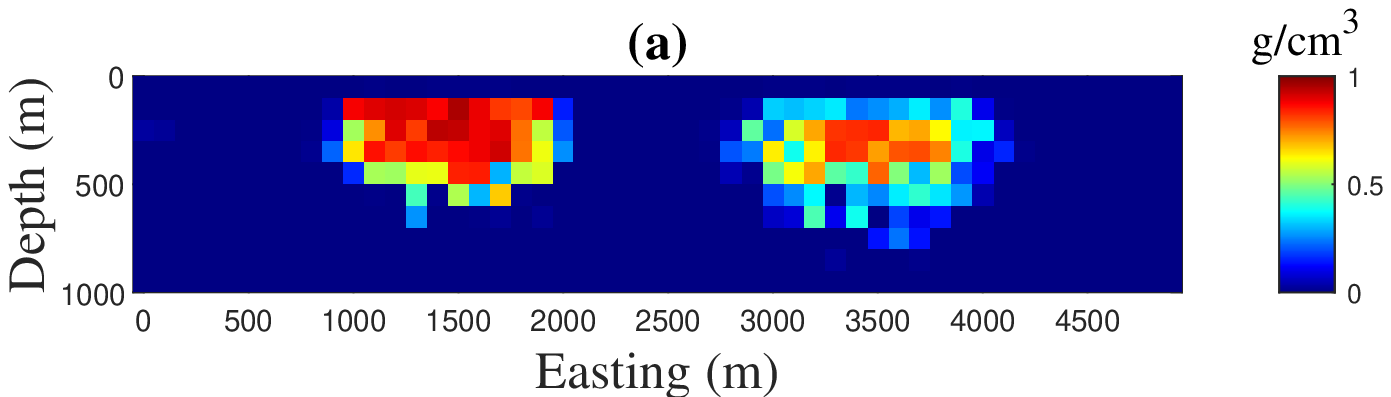}}
\subfigure{\label{9b}\includegraphics[width=.7\textwidth]{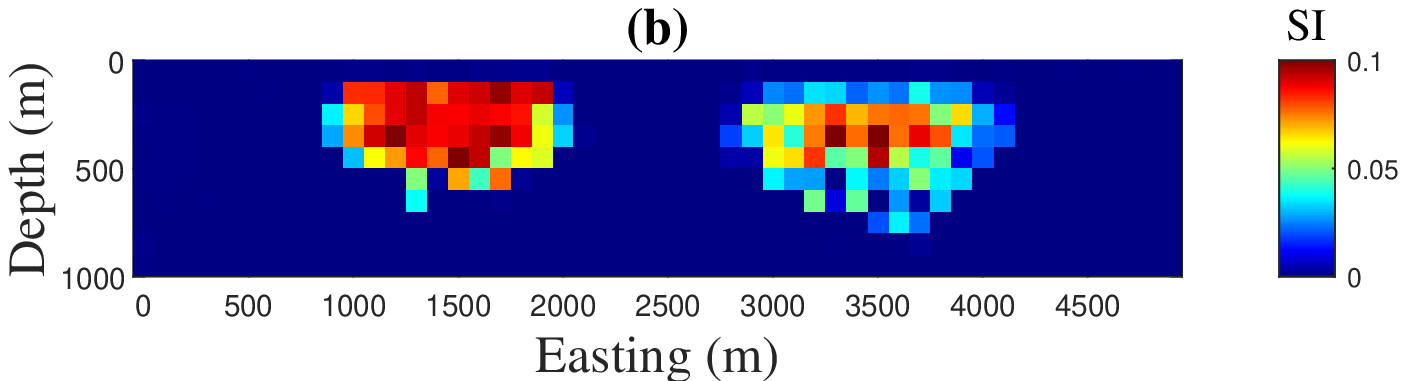}}
\caption {Cross sections of the reconstructed models for the data sets shown in~\cref{2a,2b} obtained with  $\lambda^{(1)}=1000$ and $ \lambda^{(2)}=15000$. In \cref{9a} the  density distribution, and in \cref{9b} the magnetic susceptibility distribution.} \label{fig9}
\end{figure*}

\begin{figure*}
\includegraphics[width=0.5\textwidth]{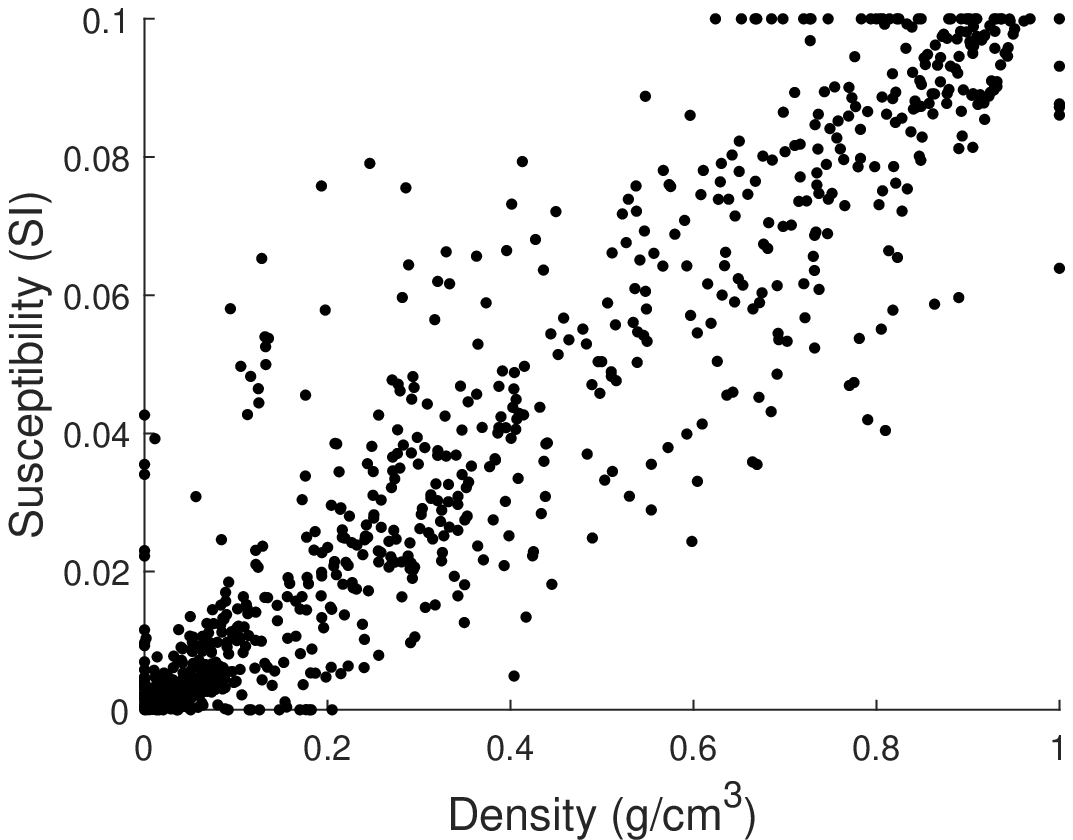}
\caption{The cross-correlation plot between the density and magnetic susceptibility model parameters for the solutions presented in~\cref{9a,9b}, obtained with $\lambda^{(1)}=1000$ and $ \lambda^{(2)}=15000$.} \label{fig10}
\end{figure*}

\begin{figure*}
\subfigure{\label{11a}\includegraphics[width=.45\textwidth]{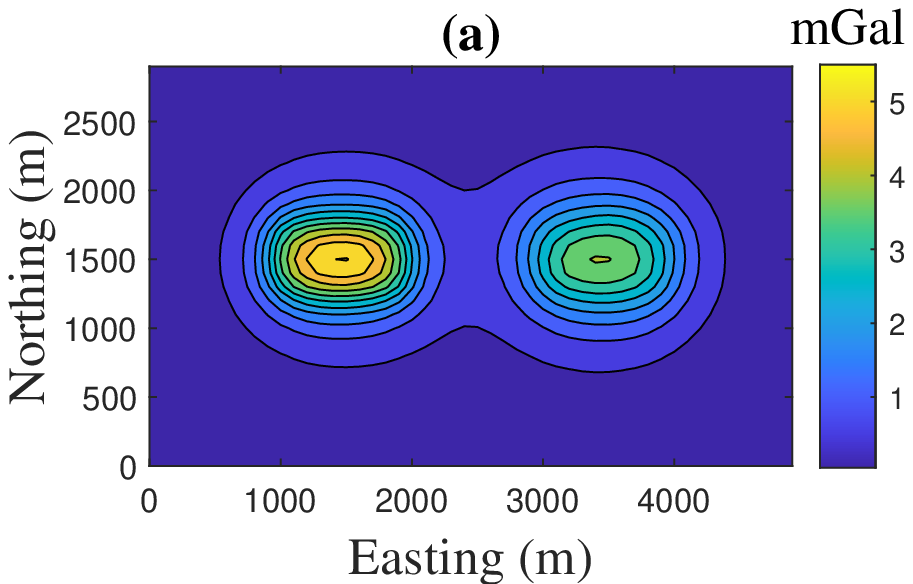}}
\subfigure{\label{11b}\includegraphics[width=.47\textwidth]{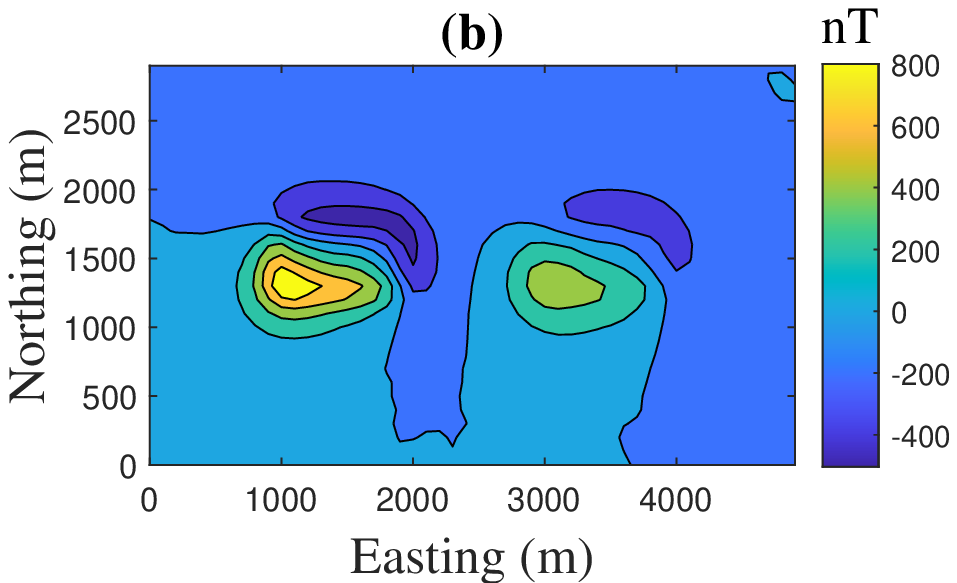}}
\caption {Data predicted by the models shown in~\cref{9a,9b}, obtained with  $\lambda^{(1)}=1000$ and $ \lambda^{(2)}=15000$. In \cref{11a} the  vertical component of the gravity, and in \cref{11b} the total magnetic field.} \label{fig11}
\end{figure*}

\subsubsection{Inversion with Gramian constraint in the space of weighted parameters: $\tilde{\lambda}^{(1)}=5000$,  and $ \tilde{\lambda}^{(2)}=10000$}\label{Gramianweightedspace}
In~\cref{steepest} we present a detailed theoretical discussion of the difference between the application of the Gramian constraint in the spaces of the original and weighted model parameters. Here, we illustrate the results when the algorithm, with Gramian in the weighted space, is applied on our synthetic example. Generally, we have found that the Lagrange parameters associated with $\tilde{S}_G$ need to be selected larger than their counterparts for $S_G$. We present the results for $\tilde{\lambda}^{(1)}=5000$,  and $ \tilde{\lambda}^{(2)}=10000$, in which the algorithm terminates after $47$ iterations. The reconstructed density and magnetic susceptibility models are illustrated in~\cref{12a,12b}, respectively. They are not as similar as we expect from the application of the Gramian constraint, as we showed in~\cref{diffLagrange}. The relative errors of the reconstructed  models are   $RE^{(1)}=0.5419$ and $RE^{(2)}=0.6870$, respectively. The cross-correlation plots for the weighted and unweighted model parameters, at the final iteration, are shown in~\cref{fig13}. This clearly indicates that the application of the Gramian constraint in the weighted space provides correlation between weighted model parameters, and not between the original unweighted parameters. Finally, the predicted gravity and magnetic data sets are illustrated in~\cref{fig14}.  

\begin{figure*}
\subfigure{\label{12a}\includegraphics[width=.7\textwidth]{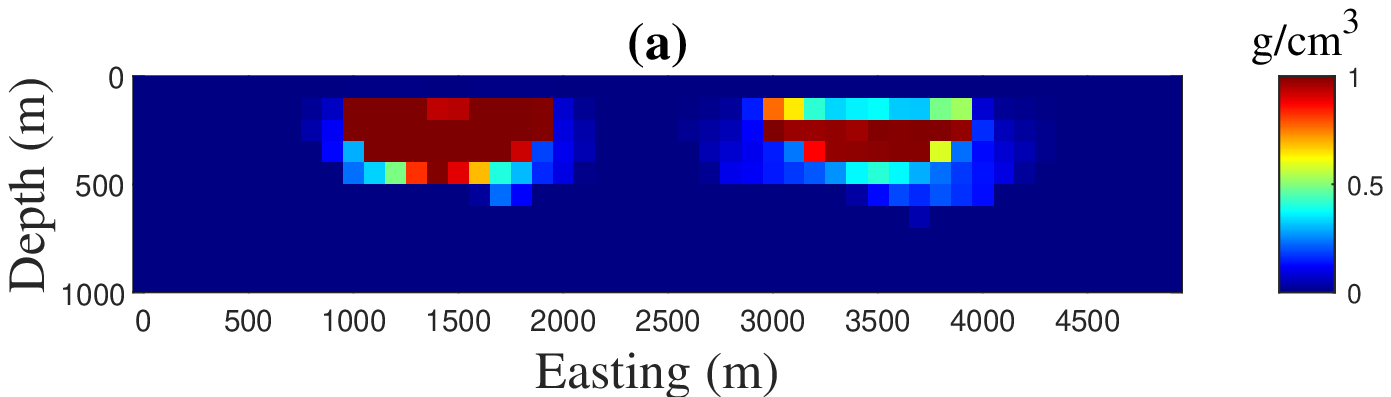}}
\subfigure{\label{12b}\includegraphics[width=.7\textwidth]{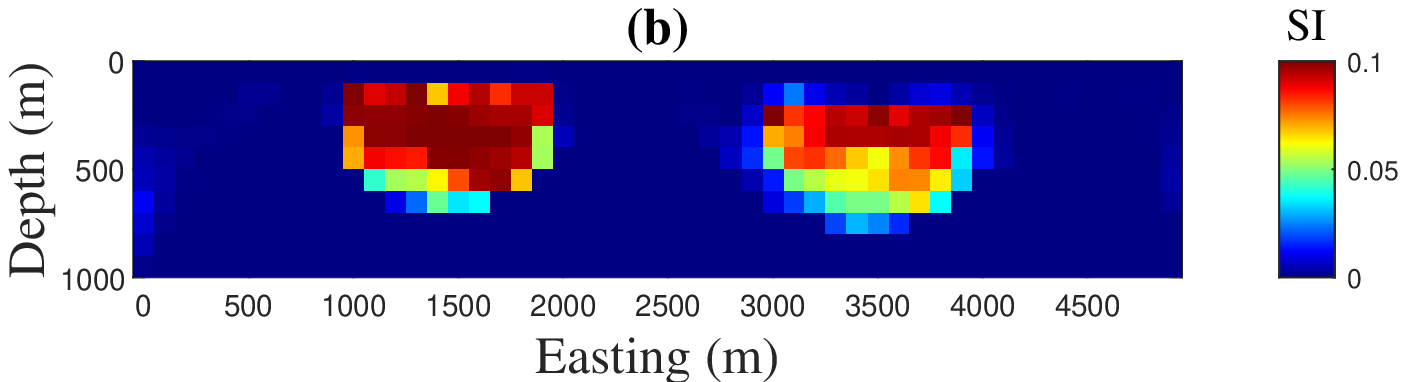}}
\caption{Cross sections of the reconstructed models for the data sets shown in~\cref{2a,2b} obtained with  the application of the Gramian constraint in the space of the weighted  model parameters with  $\tilde{\lambda}^{(1)}=5000$ and $ \tilde{\lambda}^{(2)}=10000$. In~\cref{12a} the  density distribution, and in~\cref{12b} the magnetic susceptibility distribution.} \label{fig12}
\end{figure*}

\begin{figure*}
\subfigure{\label{13a}\includegraphics[width=0.45\textwidth]{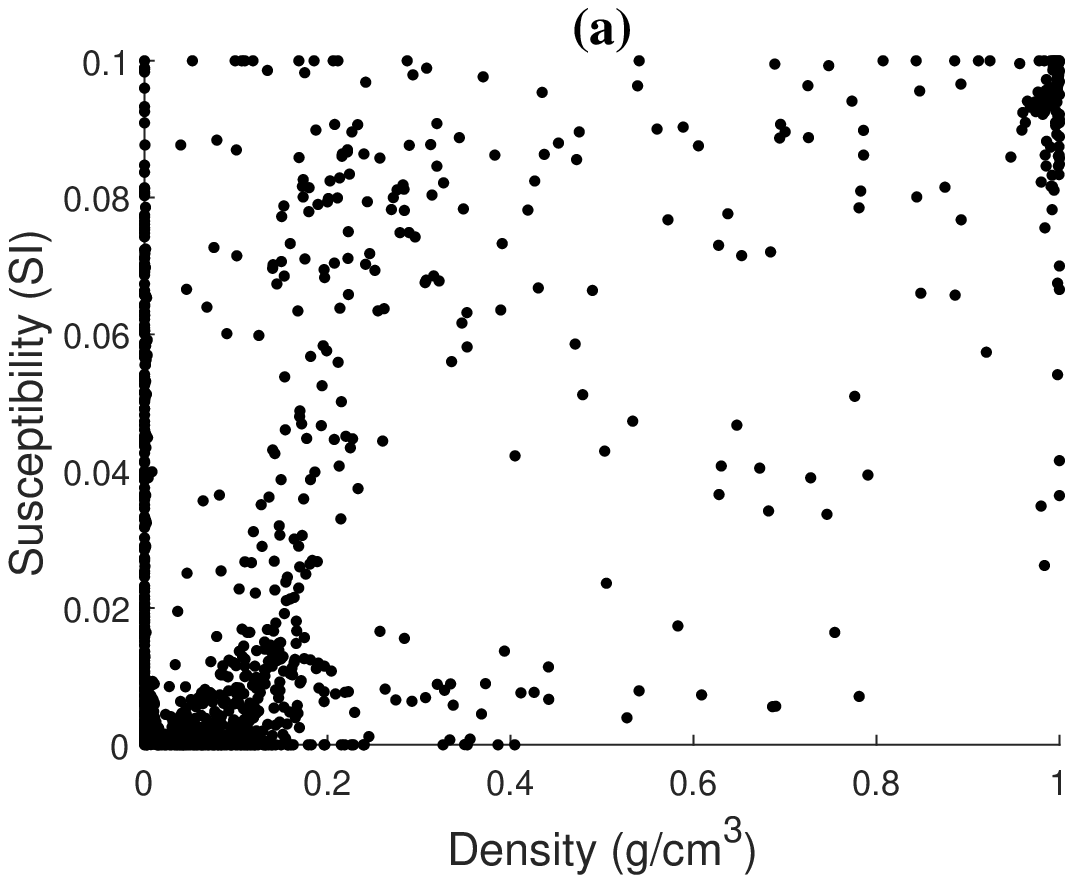}}
\subfigure{\label{13b}\includegraphics[width=0.45\textwidth]{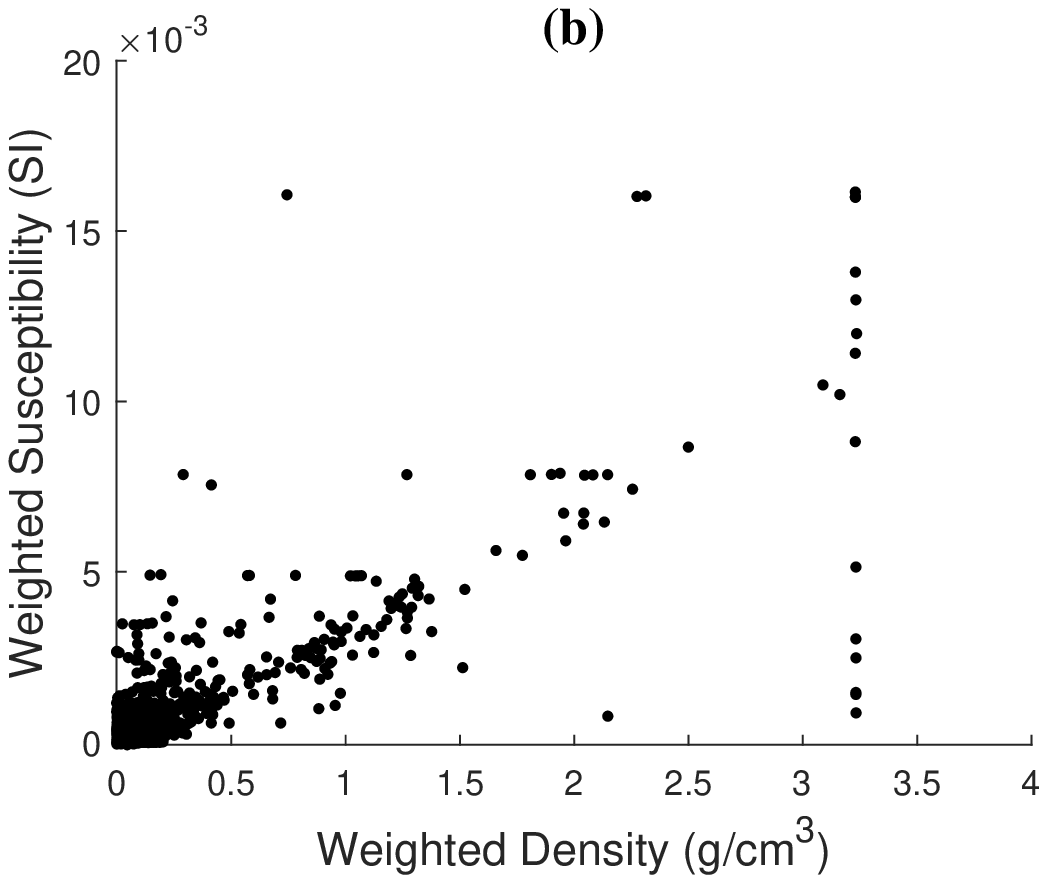}}
\caption{The cross-correlation plot between the model parameters obtained at final iteration when the Gramian constraint is applied in the space of the weighted model parameters with  $\tilde{\lambda}^{(1)}=5000$ and $\tilde{\lambda}^{(2)}=10000$. In \cref{13a} model parameters in the original space, and in \cref{13b} the weighted model parameters.} \label{fig13}
\end{figure*}

\begin{figure*}
\subfigure{\label{14a}\includegraphics[width=.45\textwidth]{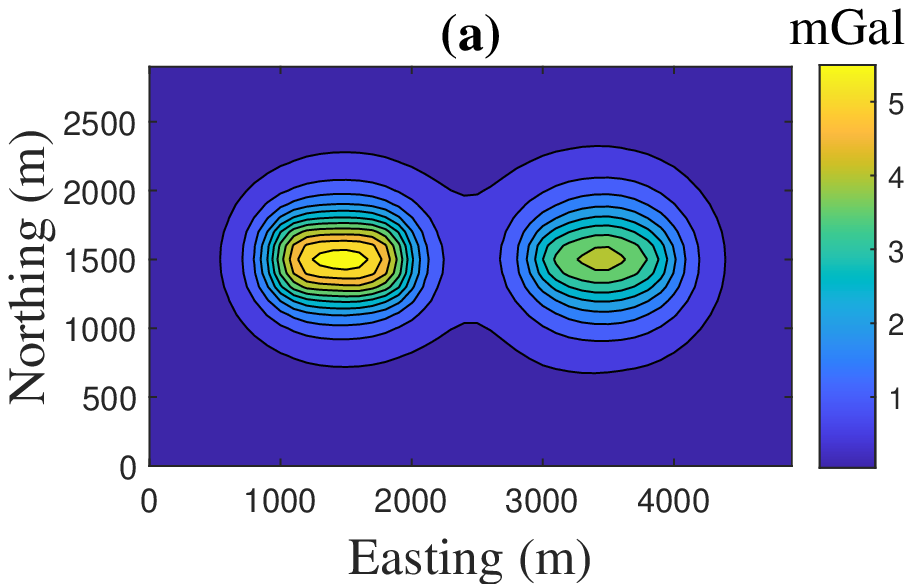}}
\subfigure{\label{14b}\includegraphics[width=.47\textwidth]{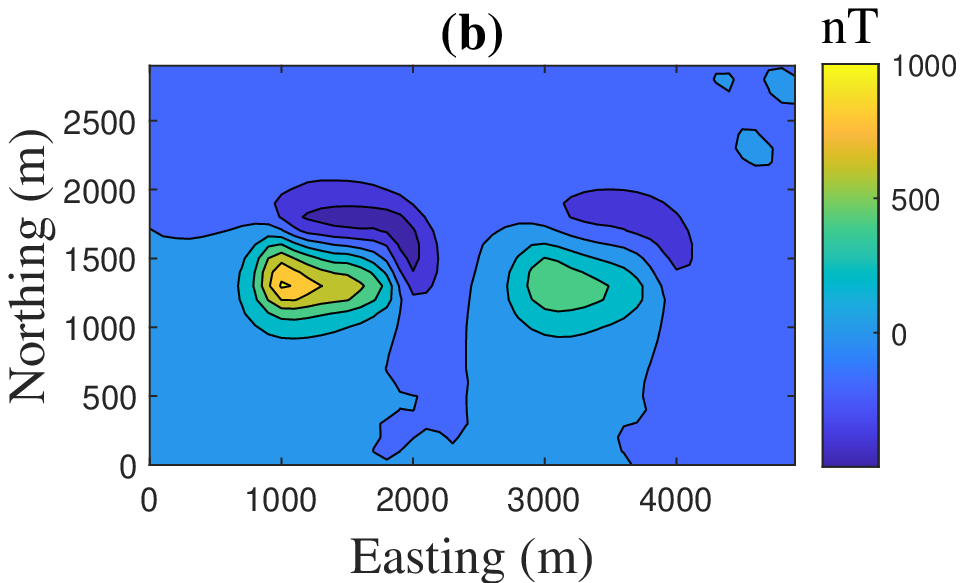}}
\caption {Data predicted by the models shown in~\cref{12a,12b}.  In \cref{14a} the vertical component of the gravity, and in \cref{14b} the total magnetic field.} \label{fig14}
\end{figure*}

\subsubsection{Joint minimum length inversion: $\WL=\bfeye$}\label{grammian:smoothing}
We now discuss, and then illustrate an example of,  the performance of the joint Gramian inversion in conjunction with the smoothing norm that is obtained using $\WL=\bfeye$, here denoted as the joint minimum length inversion. In general, regardless of the choice of $\lambda^{(i)}=0$ or $>0$,  we observe that the algorithm requires many more iterations for convergence, as compared to the focusing inversion. On the other hand, consistent with smoothing inversion, the data misfits gradually and stably decrease with the iterations, and, as has been seen in other situations, Renaut et al.  \shortcite{R:2020};  Vatankhah et al. \shortcite{VLR:2022}, the data misfit term for the magnetic problem satisfies the noise level much faster than is the case for the gravity data misfit. Consequently, we generally need both  $\lambda^{(2)}\gg \lambda^{(1)}$ and $\alpha_1^{(2)}> \alpha_1^{(1)}$ to provide an equivalent balance of the regularization and coupling for each set of model parameters when performing the joint inversion, 

We now present an illustrative example of the results that are obtained, with the parameters chosen based on the observations of the relative sizes of the parameters and the slower convergence in this context. We select $K_{\mathrm{max}}=500$, $\lambda^{(1)}=50$ and $\lambda^{(2)}=800000$, and  $\alpha_{1}^{(1)}=20000$ and  $\alpha_{1}^{(2)}=40000$, with also the lower bound $\alpha_{k}^{(i)}\ge 50$ in order to assure that the minimum length constraint is maintained throughout the iteration. After $387$ iterations, and only about $22$~s, the convergence criteria are satisfied and the inversion terminates. The reduced computational time per iteration is due to the use of the smoothing norm with $\WL=\bfeye$, replacing the iteratively-obtained $\WL$ required for focusing inversion. The reconstructed density and magnetic susceptibility models, which are  illustrated in~\cref{15a,15b},  are smooth and   similar,  the relative errors are comparable, $RE^{(1)}=0.6117$ and $RE^{(2)}=0.5960$, and, as shown in \cref{fig16}, the parameters are highly correlated. Again, the  predicted gravity and magnetic data sets, illustrated in~\cref{fig17},  are in  good agreement with the observed data at the noise level. We conclude that, although the models are geophysically acceptable  and can provide major features of the subsurface targets, the bodies are much smoother than desired for extracting the precise extents of the targets.

\begin{figure*}
\subfigure{\label{15a}\includegraphics[width=.7\textwidth]{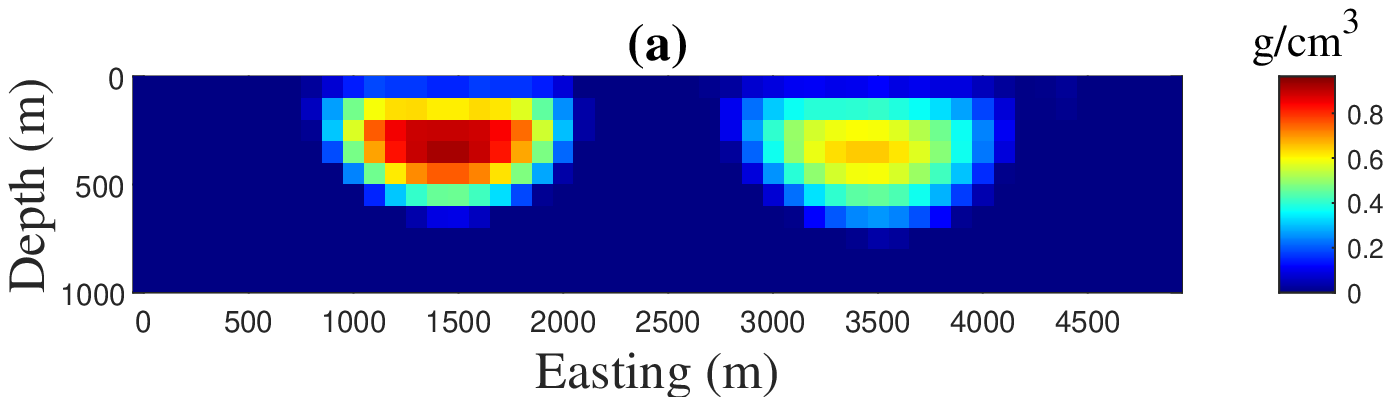}}
\subfigure{\label{15b}\includegraphics[width=.7\textwidth]{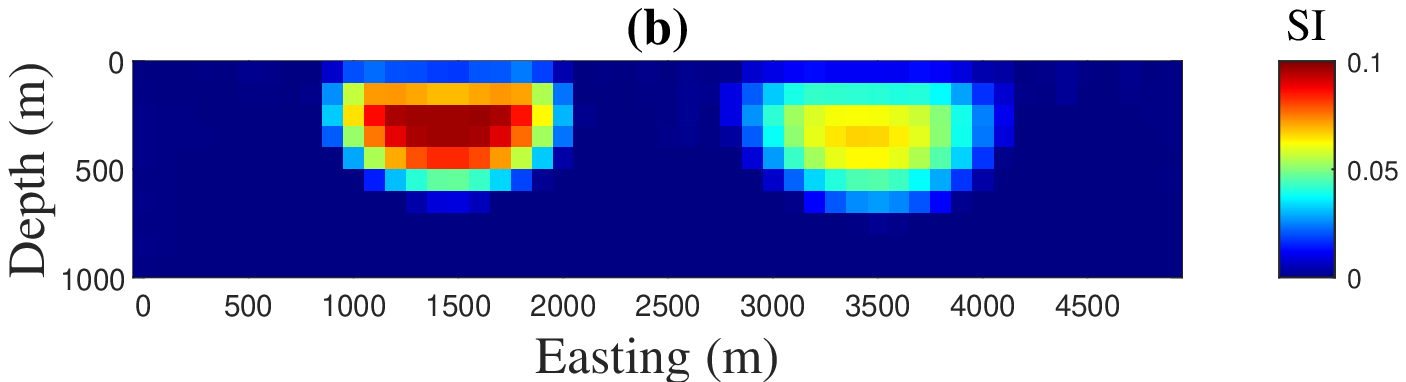}}
\caption{Cross sections of the reconstructed models for the data sets shown in~\cref{2a,2b} obtained with  $\lambda^{(1)}=50$ and $ \lambda^{(2)}=800000$, and the joint minimum length algorithm. In~\cref{15a} the  density distribution, and in~\cref{15b} the magnetic susceptibility distribution.} \label{fig15}
\end{figure*}

\begin{figure*}
\includegraphics[width=0.5\textwidth]{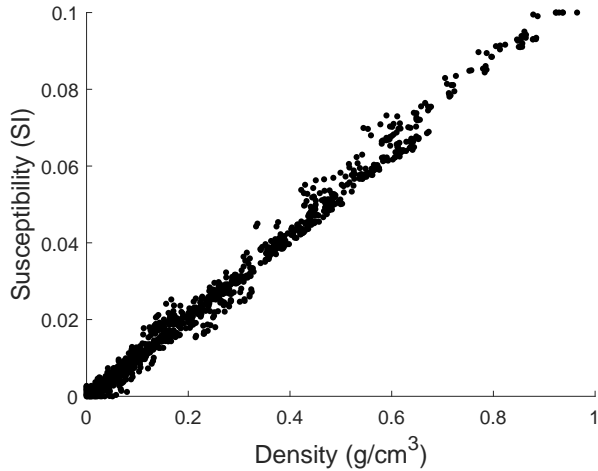}
\caption{The cross-correlation plot between the density and magnetic susceptibility model parameters for the solutions presented in~\cref{15a,15b}, obtained with  $\lambda^{(1)}=50$ and $ \lambda^{(2)}=800000$, and the joint minimum length algorithm.} \label{fig16}
\end{figure*}

\begin{figure*}
\subfigure{\label{17a}\includegraphics[width=.45\textwidth]{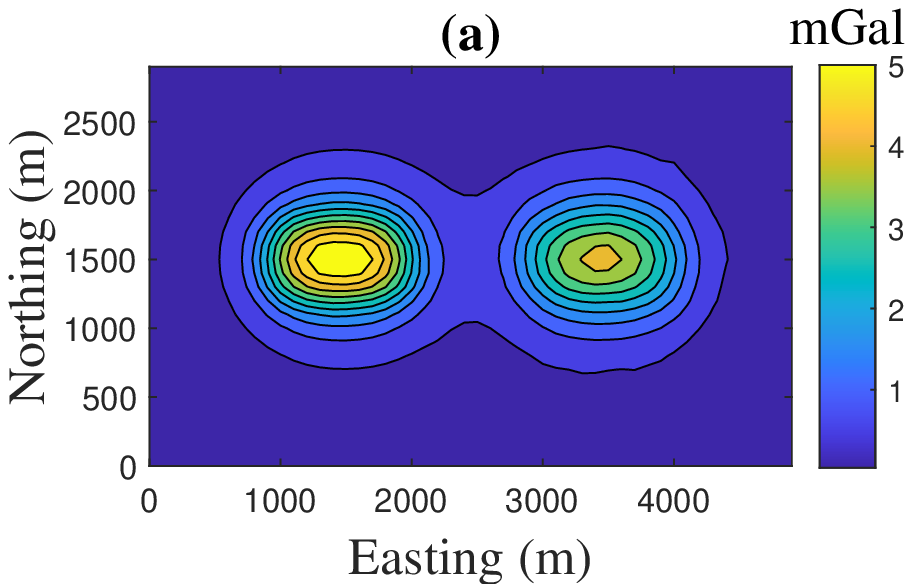}}
\subfigure{\label{17b}\includegraphics[width=.47\textwidth]{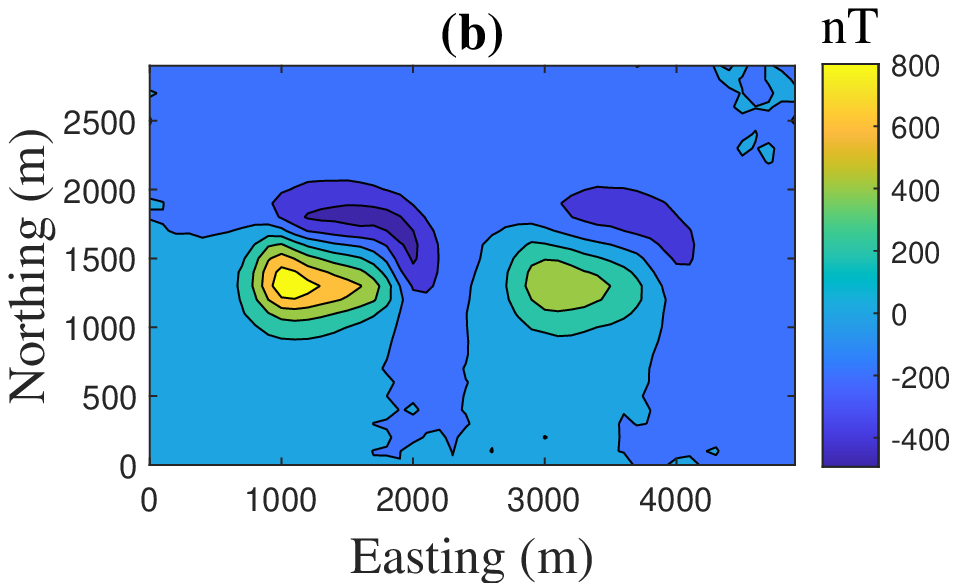}}
\caption {Data predicted by the models shown in~\cref{15a,15b}, obtained with  $\lambda^{(1)}=50$ and $ \lambda^{(2)}=800000$, and the joint minimum length algorithm. In \cref{17a} the  vertical component of the gravity, and in \cref{17b} the total magnetic field.} \label{fig17}
\end{figure*}

\subsection{Large-scale model with multiple targets:  $\lambda^{(1)}=2$ and  $\lambda^{(2)}=100$.}\label{multiple}
We now consider the inversion of large-scale data for a more complex model that consists of  five bodies with various geometries, sizes, and depths, as illustrated in the $3$D iso-surface view of the model in~\cref{fig18}. Three plane-sections of the density distribution for this model are also shown in~\cref{fig19}.  The gravity and magnetic data sets are  generated on the surface  over the $150 \times 100$  uniform grid with $100$~m grid spacing.  The resulting noise-contaminated data sets, using the noise model described in~\cref{noise}, are illustrated in~\cref{fig20}. To perform the inversion, the model region of depth $1000$~m is discretized into $ 150 \times 100 \times 10 = 150000$ prisms of size $100$~m in each dimension. Correspondingly, the sensitivity matrices $A^{(i)}$ are each of size $15000 \times 150000$, and the storage of both of these dense matrices would require least $36$~GB ($3.6 e10$ bytes) memory, which is not feasible for standard desktop or laptop computers. Fortunately, the use of the \bttb~structure in each case makes it feasible to solve the problem with a much smaller memory requirement. The transform matrices for this problem are of size $299 \times 199 \times 10$, for each gravity and magnetic problem, and the storage requirement is approximately $19$~MB, noting that each transform matrix is complex, corresponding  to a reduction of storage by a factor of approximately $1891$. 

After $96$ iterations, and  about $4872$~s, the convergence criteria are satisfied and the inversion terminates. The reconstructed density and magnetic susceptibility models,  illustrated via the plane-sections in~\cref{fig21,fig22},  are sparse,  geophysically acceptable,   similar, and consistent with the true models. Furthermore, the relative errors are comparable,   $RE^{(1)}=0.5491$ and $RE^{(2)}=0.5615$ and there is an almost linear correlation between the model parameters, as illustrated in the cross-correlation plot in~\cref{fig23}. Finally, the predicted gravity and magnetic data sets, illustrated in~\cref{fig24},  are in  good agreement with the observed data at the noise level. These results demonstrate that it is feasible to invert large data sets at a reasonable computational cost, and that the algorithm provides realistic results, even when the model is complex. The horizontal borders of the bodies are recovered  very well, but, at depth the reconstructed models are not perfect matches for the original models. These conclusions are consistent with the results obtained for the less complex and smaller problem.

\begin{figure*}
\includegraphics[width=0.8\textwidth]{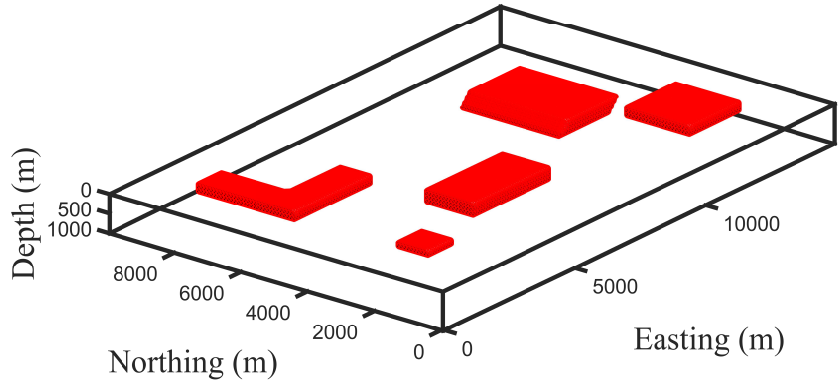}
\caption{Complex model consisting of five bodies with different shapes, dimension, and depths. The density contrast and the magnetic susceptibility of the bodies are selected as $1$~g~cm$^{-3}$ and $0.1$ SI units, respectively, embedded in a homogeneous non-susceptible background.} \label{fig18}
\end{figure*}

\begin{figure*}
\subfigure{\label{19a}\includegraphics[width=.45\textwidth]{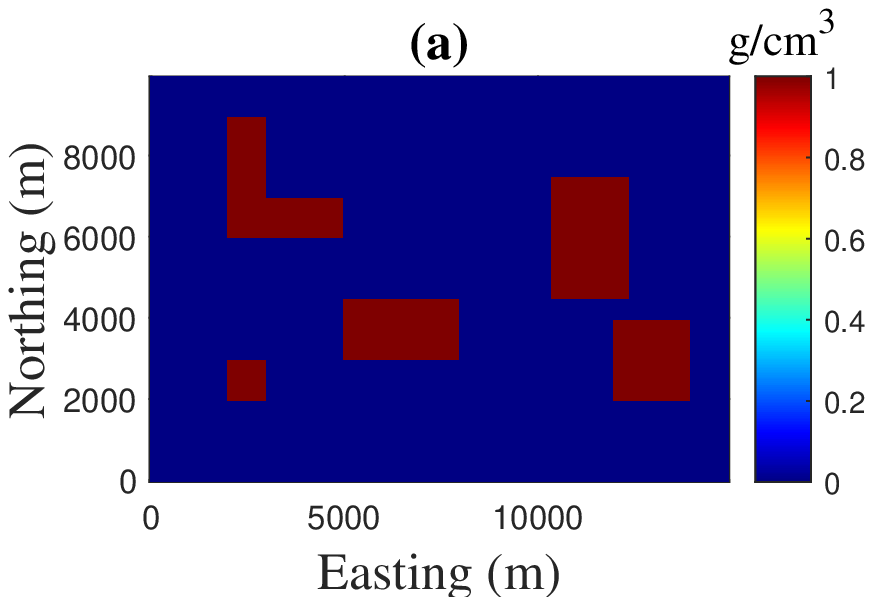}}
\subfigure{\label{19b}\includegraphics[width=.45\textwidth]{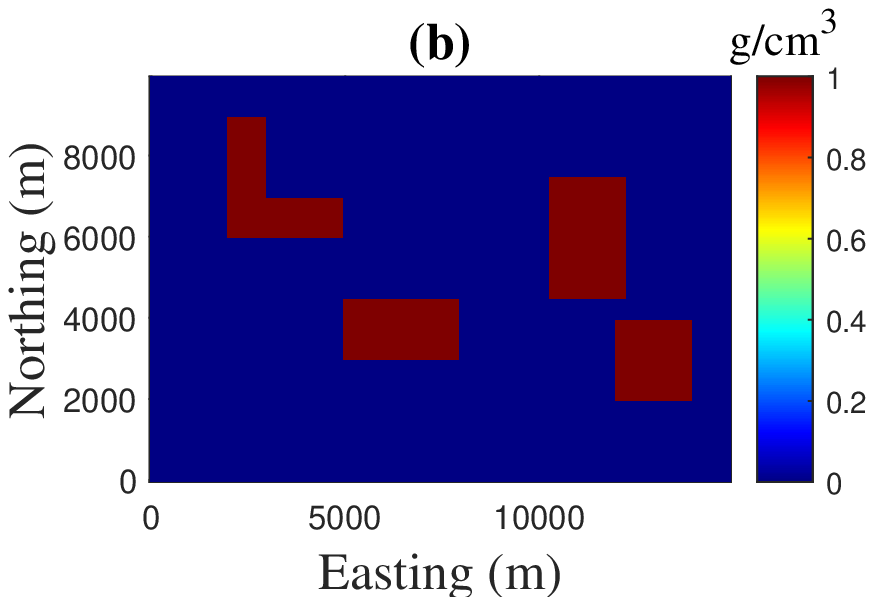}}
\subfigure{\label{19c}\includegraphics[width=.45\textwidth]{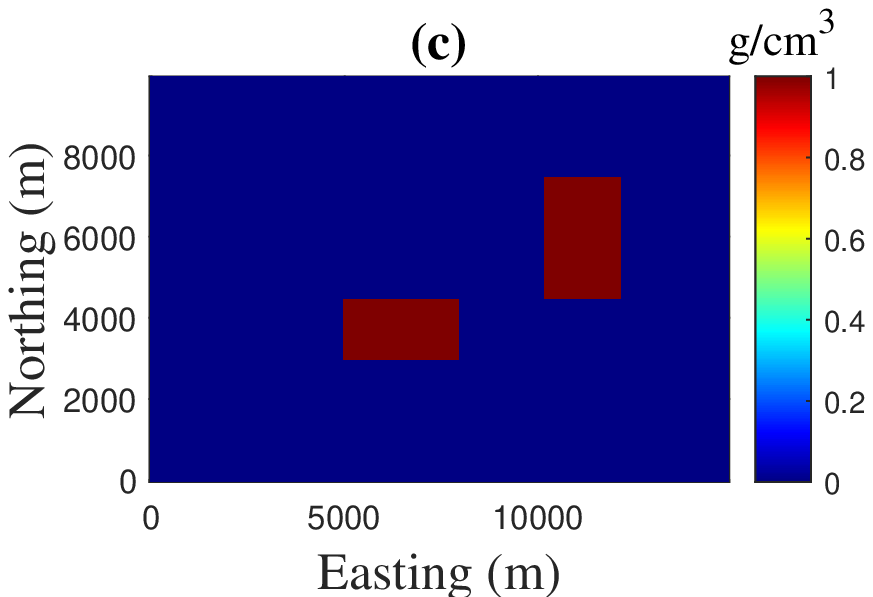}}
\caption{Three plane-sections of the model shown in~\cref{fig18}. Here, the density distribution of the model is presented at depths: in~\cref{19a} $200$~m; in~\cref{19b} $300$~m; and in~\cref{19c} $400$~m.} \label{fig19}
\end{figure*}

\begin{figure*}
\subfigure{\label{20a}\includegraphics[width=0.45\textwidth]{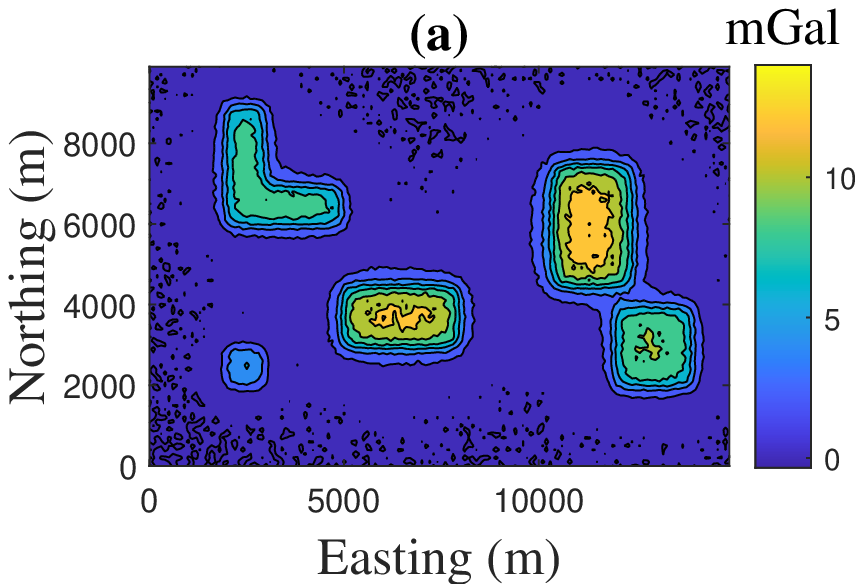}}
\subfigure{\label{20b}\includegraphics[width=0.47\textwidth]{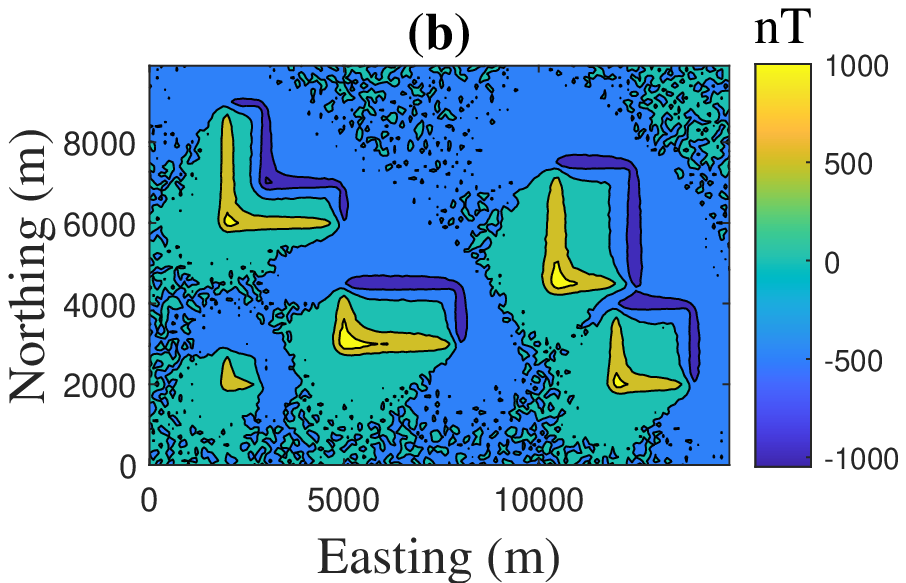}}
\caption{Noise-contaminated data produced by the model shown in~\cref{fig18}. In~\cref{20a} the vertical component of the gravity field,  and in \cref{20b} the total magnetic field.} \label{fig20}
\end{figure*}

\begin{figure*}
\subfigure{\label{21a}\includegraphics[width=.45\textwidth]{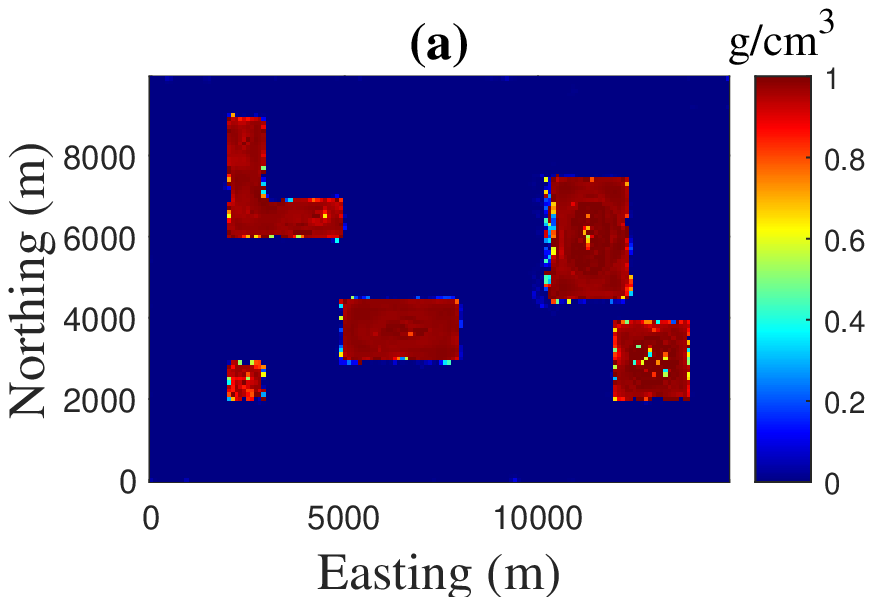}}
\subfigure{\label{21b}\includegraphics[width=.45\textwidth]{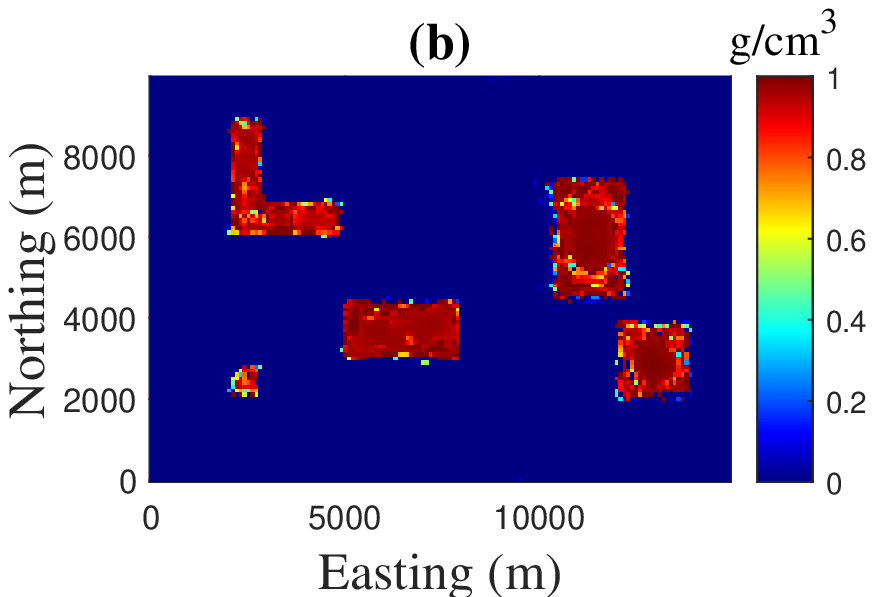}}
\subfigure{\label{21c}\includegraphics[width=.45\textwidth]{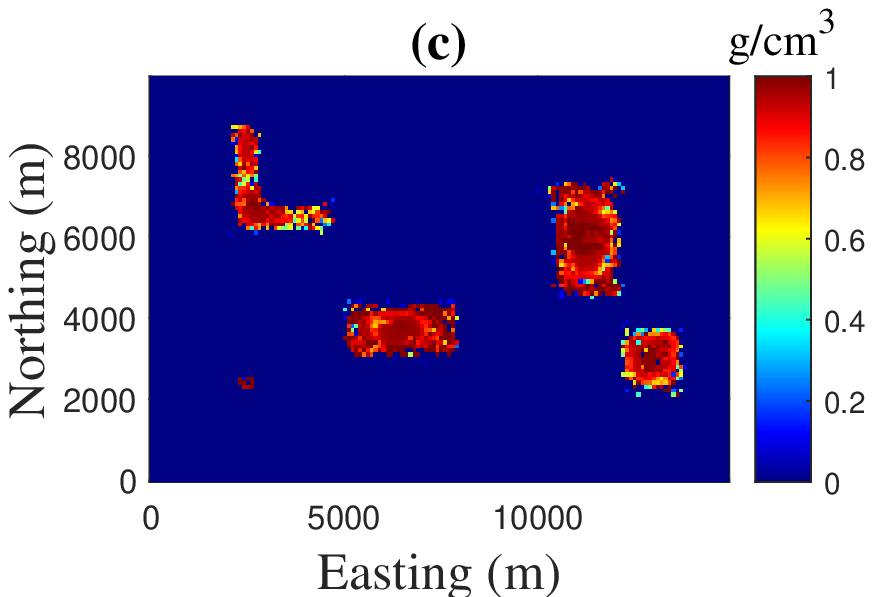}}
\caption{Plan-sections of the reconstructed density model using the presented joint inversion algorithm with Lagrange parameters, $\lambda^{(1)}=2$, $\lambda^{(2)}=100$. Consistent with \cref{fig19}, the density distribution of the model is presented at depths: in~\cref{21a} $200$~m; in~\cref{21b} $300$~m; and in~\cref{21c} $400$~m.} \label{fig21}
\end{figure*}

\begin{figure*}
\subfigure{\label{22a}\includegraphics[width=.45\textwidth]{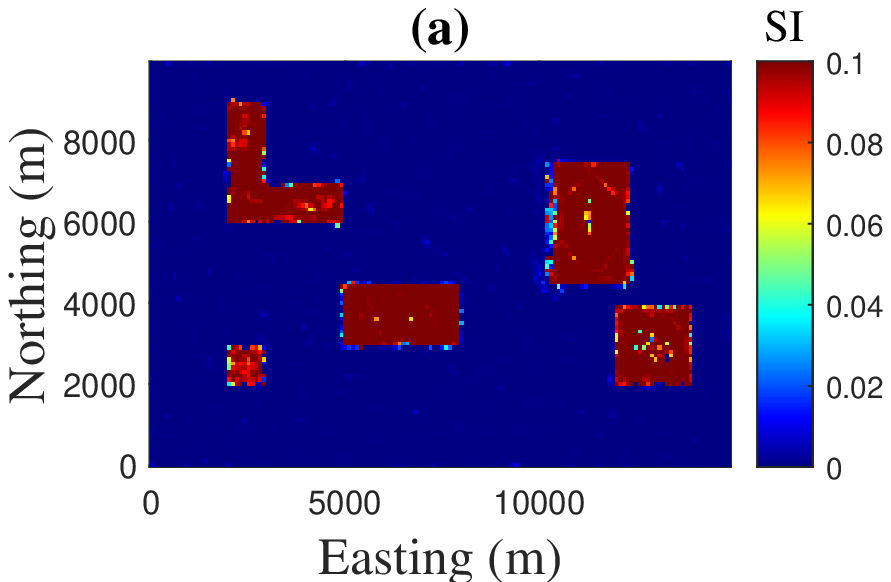}}
\subfigure{\label{22b}\includegraphics[width=.45\textwidth]{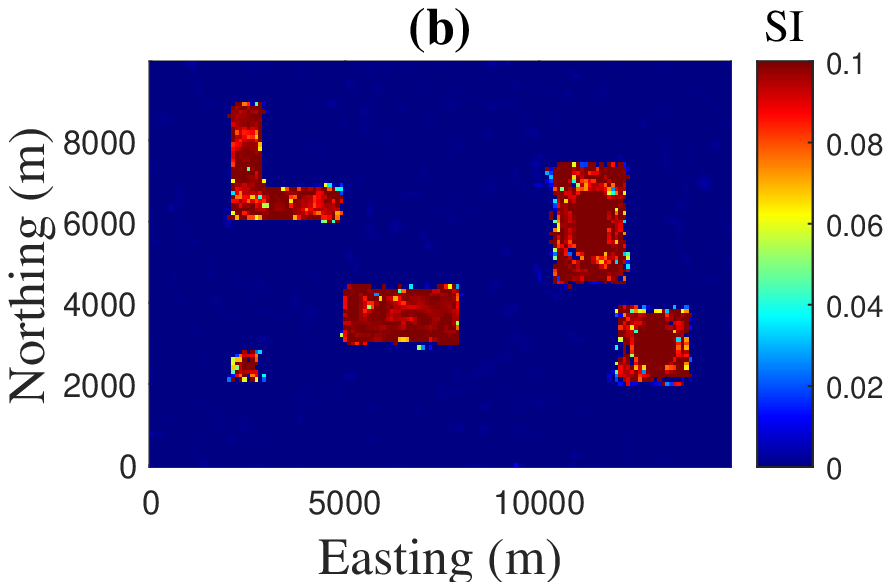}}
\subfigure{\label{22c}\includegraphics[width=.45\textwidth]{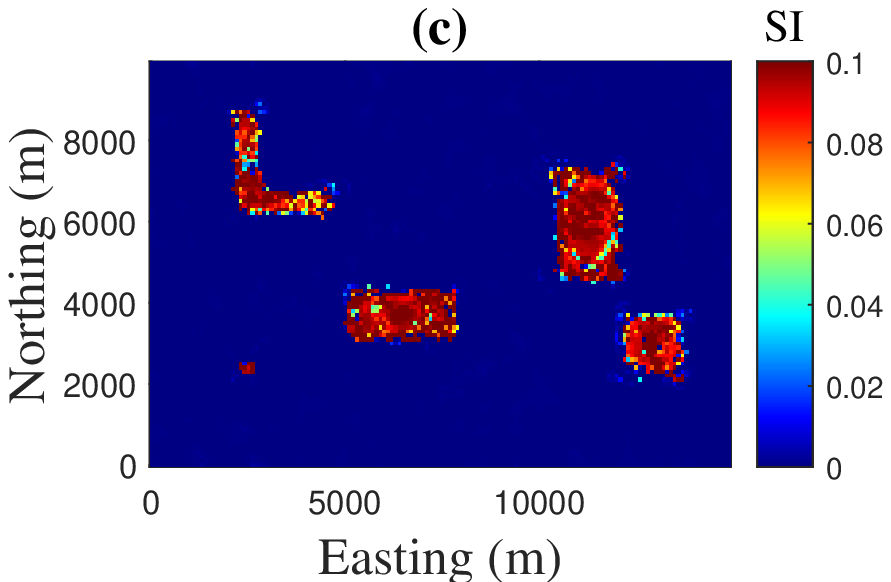}}
\caption{Plan-sections of the reconstructed magnetic susceptibility model using the presented joint inversion algorithm with Lagrange parameters $\lambda^{(1)}=2$, $\lambda^{(2)}=100$. Consistent with \cref{fig19}, the susceptibility distribution of the model is presented at depths: in~\cref{22a} $200$~m; in~\cref{22b} $300$~m; and in~\cref{22c} $400$~m.} \label{fig22}
\end{figure*}

\begin{figure*}
\includegraphics[width=.5\textwidth]{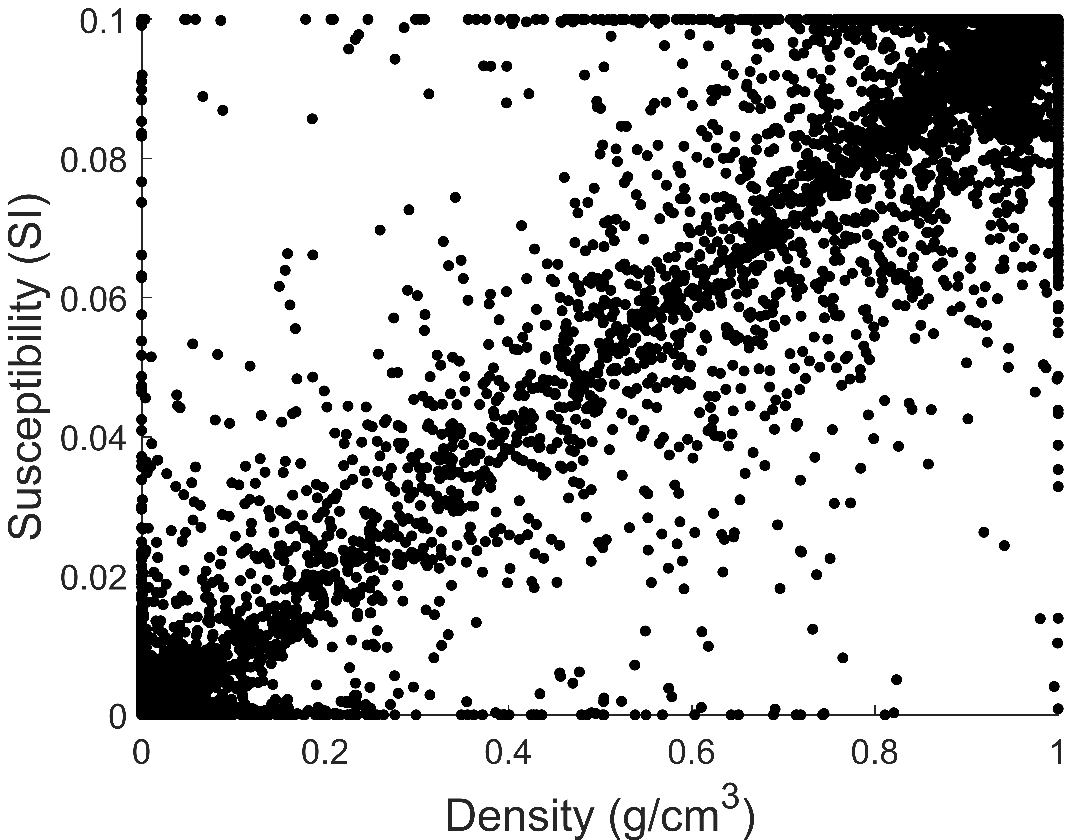}
\caption{The cross-correlation plot between the density and magnetic susceptibility model parameters for the solutions presented in~\cref{fig21,fig22}, obtained with  $\lambda^{(1)}=2$, $\lambda^{(2)}=100$.} \label{fig23}
\end{figure*}

\begin{figure*}
\subfigure{\label{24a}\includegraphics[width=.45\textwidth]{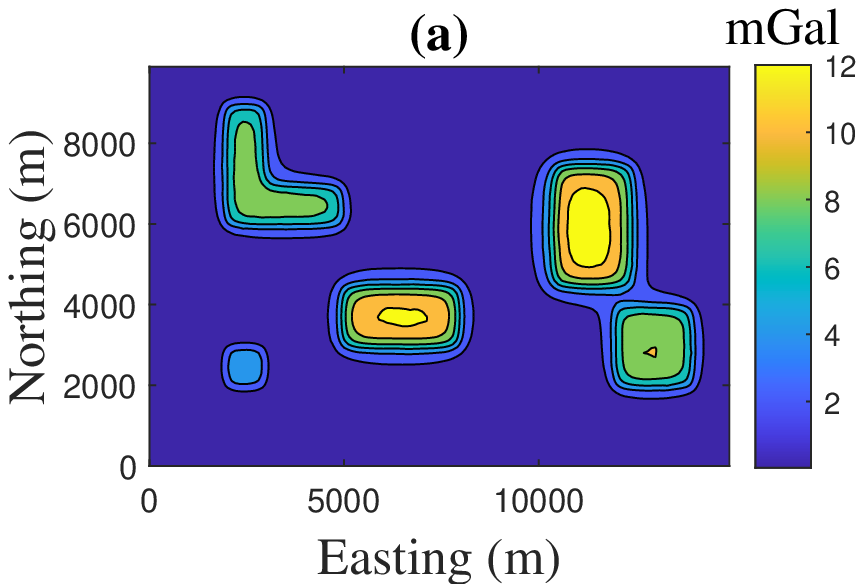}}
\subfigure{\label{24b}\includegraphics[width=.47\textwidth]{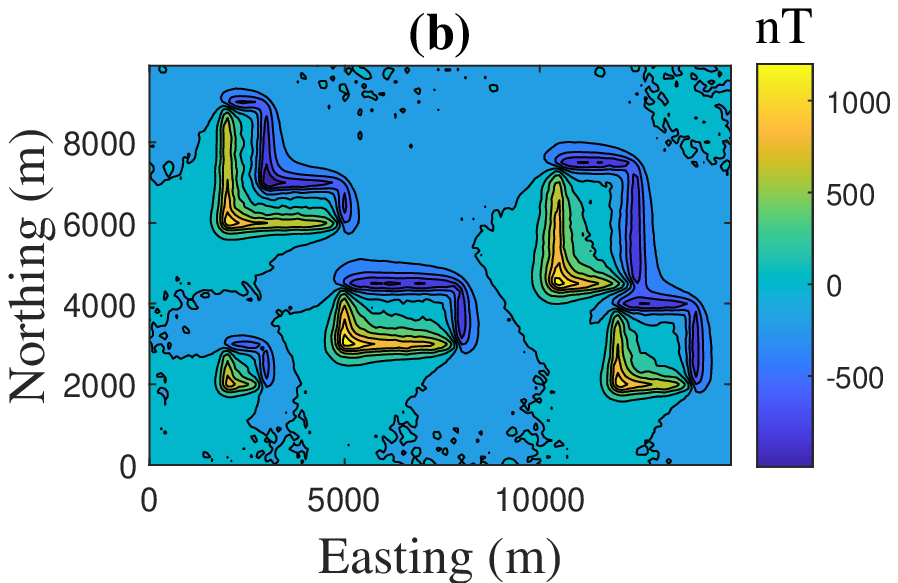}}
\caption{Data predicted by the models shown in~\cref{fig21,fig22}, obtained with  $\lambda^{(1)}=2$, $\lambda^{(2)}=100$. In~\cref{24a} the  vertical component of the gravity, and in~\cref{24b} the total magnetic field.} \label{fig24}
\end{figure*}

\section{Real data}\label{real}

 The joint inversion algorithm will be applied on gravity and aeromagnetic data obtained over the northwestern portion of the Mesoproterozoic St. Francois Terrane in southeast Missouri, USA. The geological map of the rocks in the area is shown in \cref{fig25}  \cite{IM:19}. This terrane consists of two large scale granite-rhyolite provinces, the Eastern Granite Rhyolite Province (EGRP) and the Southern Granite Rhyolite Province (SGRP) with the SGRP intruding the EGRP rocks \cite{Thomas:12}. The lithologies include high-silica rhyolitic ignimbrites, rhyolitic and trachytic flows, and granites with minor amounts of intermediate and mafic volcanic rocks \cite{Kis:81}. These volcanic rocks were formed by two major episodes of igneous activity in the Mesoproterozoic and include caldera formation and collapse and the formation of batholiths \cite{Vanschmus:96}. These igneous rocks are exposed to the southeast of the study area and covered by 1-2 km of Paleozoic limestones and sandstones  \cite{Kis:81}. The knowledge of the igneous rocks in the study area is based on scattered drillholes, deep magnetite mines and magnetic data interpretation \cite{Kis:81}.

Of interest within the study area are several iron oxide hematite–magnetite deposits (\cref{fig25}). These deposits include rare earth elements (REE), iron oxide–apatite type and iron oxide–copper–gold (IOCG) type mineralization that occur near the nonconformable surface of the Mesoproterozoic igneous basement \cite{IM:19}. Many of the deposits are thought to be related to the volcanic and igneous processes that formed the calderas within the St. Francois Terrane \cite{Kis:81}. Of major importance to our study is the Pea Ridge magnetite deposit which, however, is the only deposit that contains significant amounts of known REE elements \cite{Day:16}. The deposit contains a near-vertical dike-like body of magnetite containing four breccia pipes that are rich in REEs surrounded by rhyolitic pyroclastic rocks. The Pea Ridge breccia pipes contains  high-yield REEs adjacent to the main ore body $700$m below the surface \cite{Long:2010}. More detailed information of the Mesoproterozoic lithologies and the mineral deposits of the study area can be found in Ives $\&$ Mickus \shortcite{IM:19} and references therein.

\begin{figure*}
\includegraphics[width=.6\textwidth]{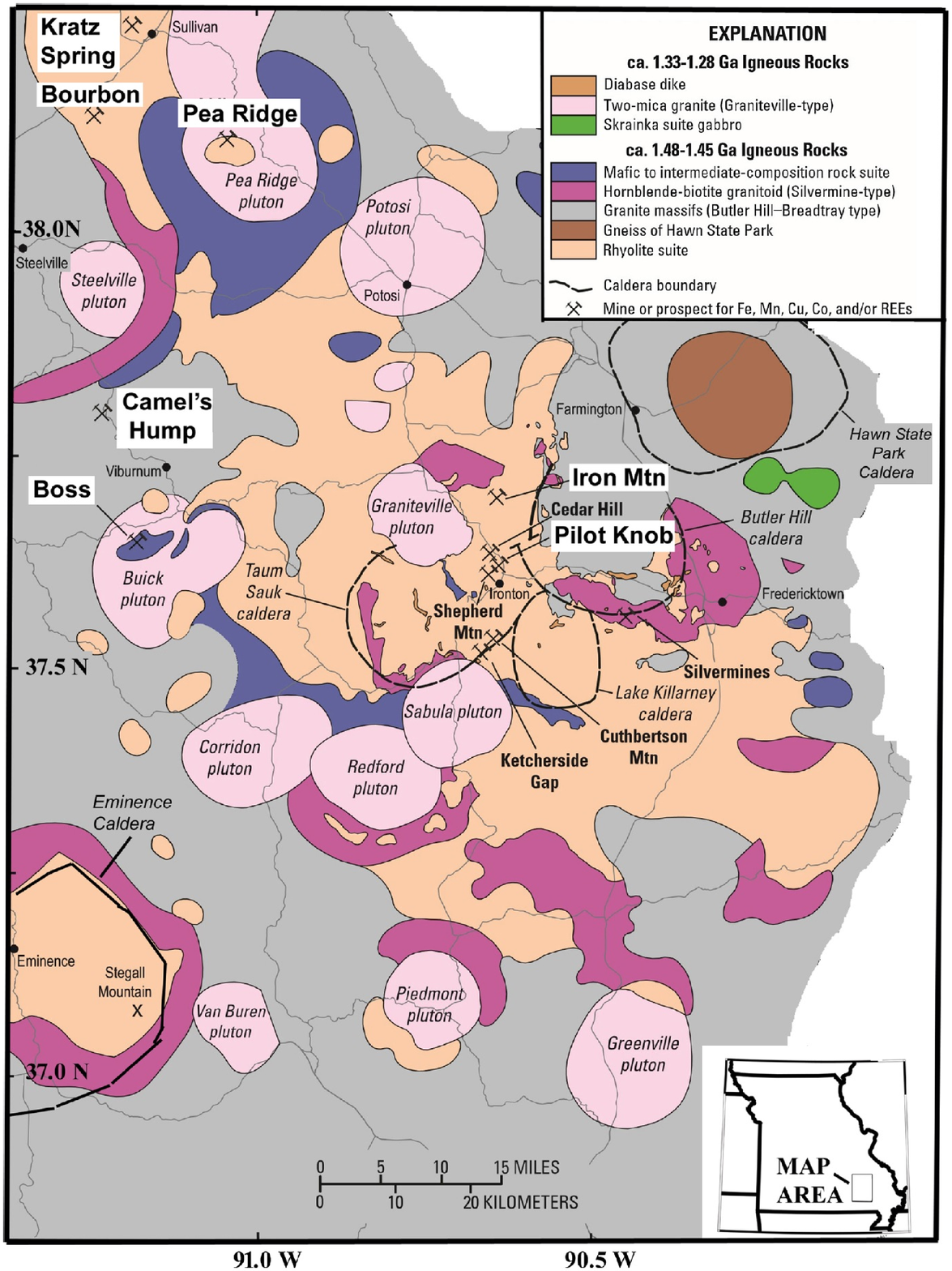}
\caption{Geologic map of the Mesoproterozoic rocks in the St. Francois Mountains Terrane, southeast of Missouri, USA. The location of the volcanic and intrusive rocks, caldera boundaries, and mineral deposits are shown. The study area, including Pea Ridge and Bourbon ore deposits, is located in the northwest section of the map.} \label{fig25}
\end{figure*}

The residual gravity and magnetic data over the survey area are shown in~\cref{26a,26b}, respectively. The data sets were gridded at $250$~m spacing to produce a grid of $146 \times 148=21608$ data points. The residual anomaly maps present the combined effects of all the upper crustal density and/or magnetization sources. The residual magnetic anomaly map clearly shows maximum anomalies over the  Bourbon and Pea Ridge mineral deposits. However since these iron ore deposits are associated with large deposits of hematite and magnetite, one would think they would be all associated with gravity maxima. The mineral deposit at Bourbon does have a positive gravity anomaly while the REE-rich Pea Ridge mineral deposit is associated with a gravity minimum. Two and one-half dimensional (2.5D) forward  modeling, Ives $\&$ Mickus \shortcite{IM:19}, showed that the main high-magnetic susceptibility ore body at Pea Ridge extends to a depth of approximately $1.5$~km with a low-density body that extends to $5$~km that was  used to explain the negative residual gravity anomaly. The cause of this low-density body is unknown but it might be associated with hydrothermal leaching of the Mesoproteozoic granitc rocks that formed the Pea Ridge deposit \cite{IM:19}.

\begin{figure*}
\subfigure{\label{26a}\includegraphics[width=.45\textwidth]{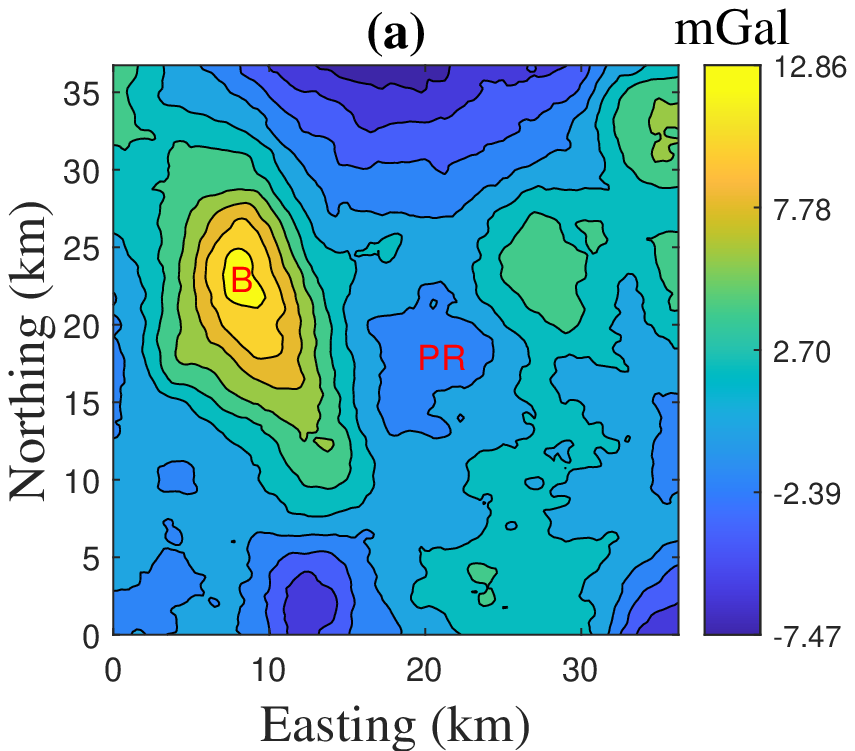}}
\subfigure{\label{26b}\includegraphics[width=.46\textwidth]{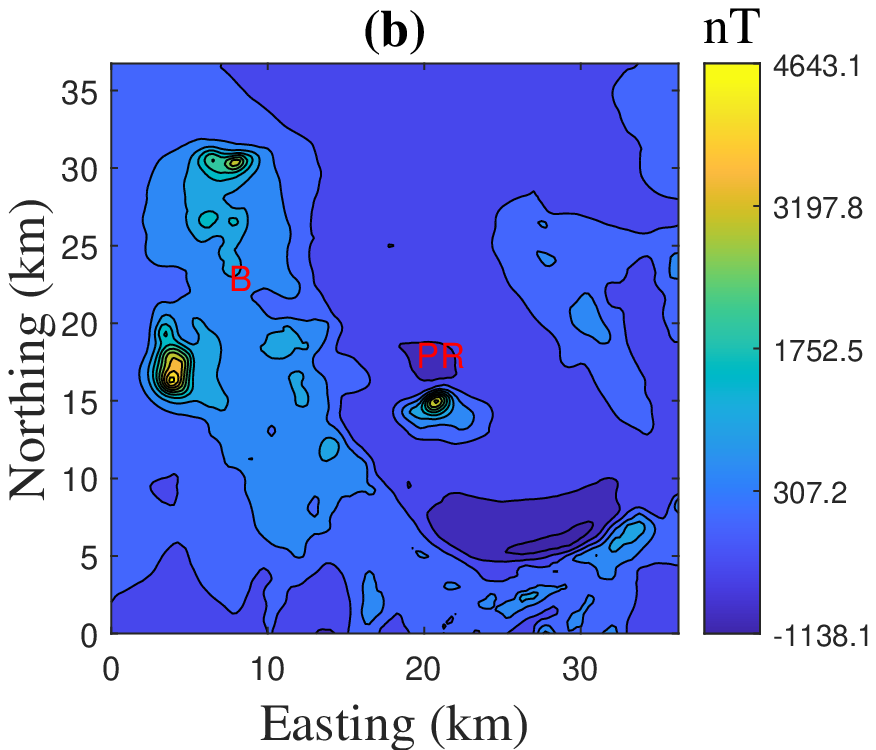}}
\caption{Residual gravity, \cref{26a},  and residual magnetic, \cref{26b},  anomalies over an area in the northwest portion of the Mesoproterozoic St. Francois Terrane. Also shown are the locations of major Fe oxide deposits including Bourbon (B) and Pea Ridge (PR).} \label{fig26}
\end{figure*}

To perform the inversion, the subsurface was divided into $146 \times 148 \times 20=432160$  rectangular prisms. The dimensions of the prisms  are $250$~m by $250$~m in the east and north directions and $500$~m in the depth direction. The model has $20$ layers and thus extends to a depth of $10$~km. We assume that the gravity and magnetic data are contaminated with noise, according the noise model in~\cref{noise}, with the pairs $(\tau_1, \tau_2)$  selected as $(0.01, 0.025)$ and $(0.01, 0.020)$ for the gravity and magnetic problems, respectively. The intensity of the geomagnetic field, the inclination, and the declination are  $53950$ nT, $67.3^{\circ}$, and $0.87^{\circ}$, respectively. Based on \textit{a priori} information, Ives $\&$ Mickus \shortcite{IM:19}, upper and lower bounds, $-0.2\leq \bfm^{(1)}\leq 0.3$~g~cm$^{-3}$ and $ 0 \leq \bfm^{(2)} \leq 0.2$,  in   SI units,  are imposed. The initial density and susceptibility models are selected as $\bfma^{(i)} = \bfzero$, $i=1$, $2$, and the maximum number of iterations, $K_{\mathrm{max}}$, is fixed as $200$.

The initial parameters $\alpha_{1}^{(1)}$ and $\alpha_{1}^{(2)}$ are set to  $20000$. Here, we also determine lower bounds $\alpha_{k}^{(1)}\ge 5$  and $\alpha_{k}^{(2)}\ge 20$. The algorithm is implemented for Lagrange parameters $\lambda^{(1)}=2$, and $\lambda^{(2)}=5$. The inversion algorithm terminates after $186$ iterations. The required time is approximately $20$ hours. We note that the problem is large and we use an ordinary laptop computer. It is the ability of the presented algorithm which allows us to perform such large-scale joint inversion problem. Three cross-sections of the reconstructed density and magnetic susceptibility models, over major subsurface targets, are illustrated in~\cref{fig27,fig29}. Additionally, three depth sections of the models are also shown in~\cref{fig28,fig30}. The reconstructed density and magnetic susceptibility models are approximately similar. The models, generally, illustrate that the igneous bodies in the area reside in the upper crust, extending to a maximum depth of $5$ to $6$~km, except for the north of the Bourbon mineral deposit in which the subsurface target is deeper. For the Pea Ridge deposit, the reconstructed models indicate a positive density and magnetic susceptibility body, that extends from $1$~km to approximately $2$~km, above a low density region. The extension of the low density region in depth approximately agrees with that proposed by Ives \& Mickus \shortcite{IM:19}. In the western section of the model, there are deeper high density and magnetic susceptibility bodies. The cross-correlation plot of the reconstructed density and magnetic susceptibility model parameters is shown in~\cref{fig31}. Finally, the predicted gravity and magnetic data sets are illustrated  in~\cref{fig32}.

\begin{figure*}
\subfigure{\label{27a}\includegraphics[width=.48\textwidth]{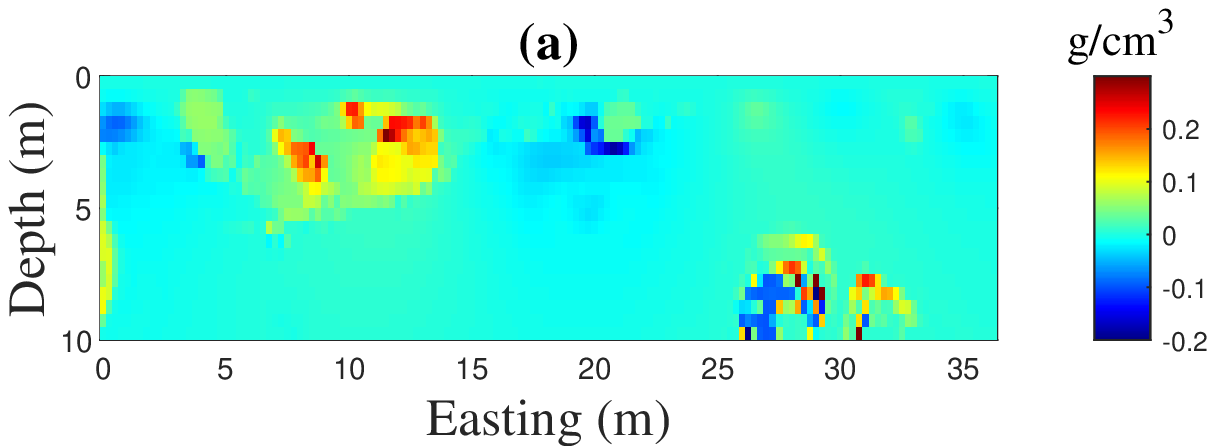}}
\subfigure{\label{27b}\includegraphics[width=.48\textwidth]{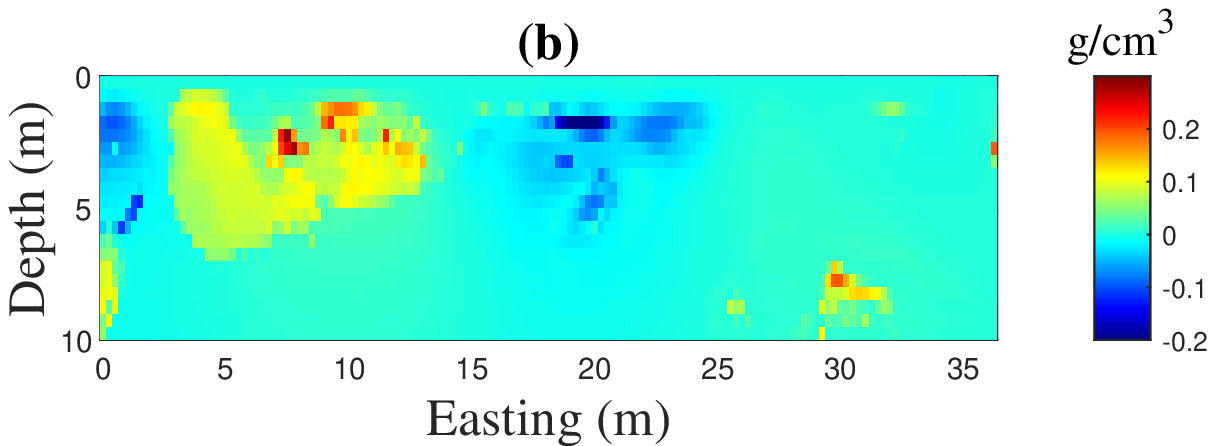}}
\subfigure{\label{27c}\includegraphics[width=.48\textwidth]{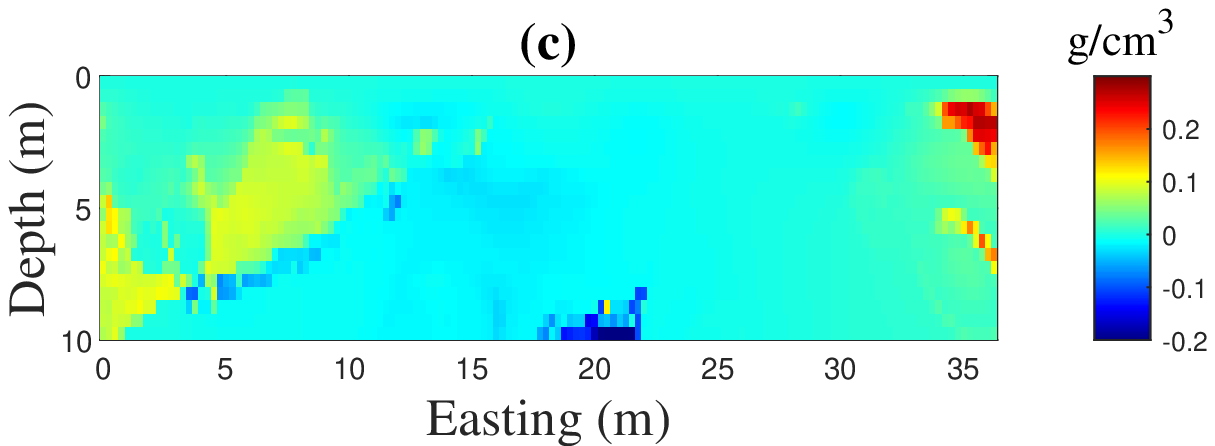}}
\caption{Cross-sections of the reconstructed density model using the joint inversion algorithm applied on the data sets shown in~\cref{26a,26b} with Lagrange parameters, $\lambda^{(1)}=2$, $\lambda^{(2)}=5$. The sections are at: in~\cref{27a} $16$~km Northing (over Pea Ridge); in~\cref{27b} $18$~km Northing; and in~\cref{27c} $30$~km Northing.} \label{fig27}
\end{figure*}

\begin{figure*}
\subfigure{\label{28a}\includegraphics[width=.44\textwidth]{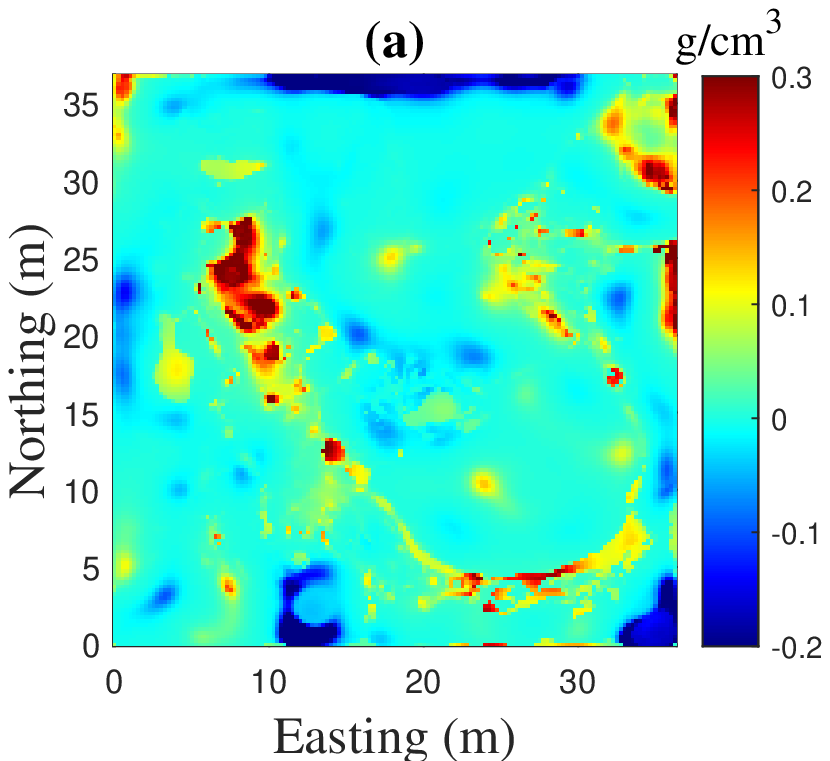}}
\subfigure{\label{28b}\includegraphics[width=.44\textwidth]{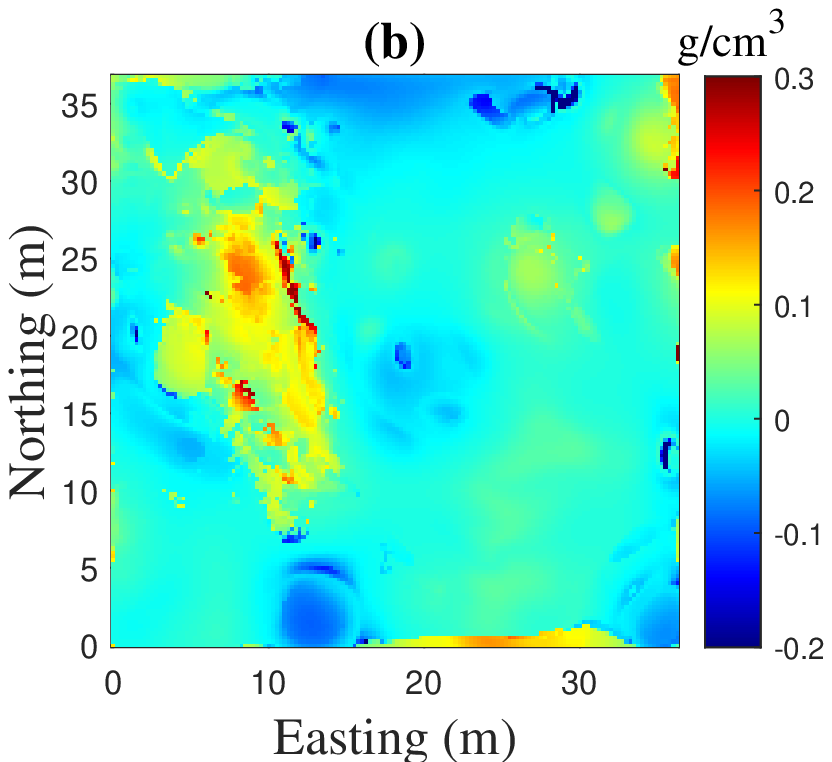}}
\subfigure{\label{28c}\includegraphics[width=.44\textwidth]{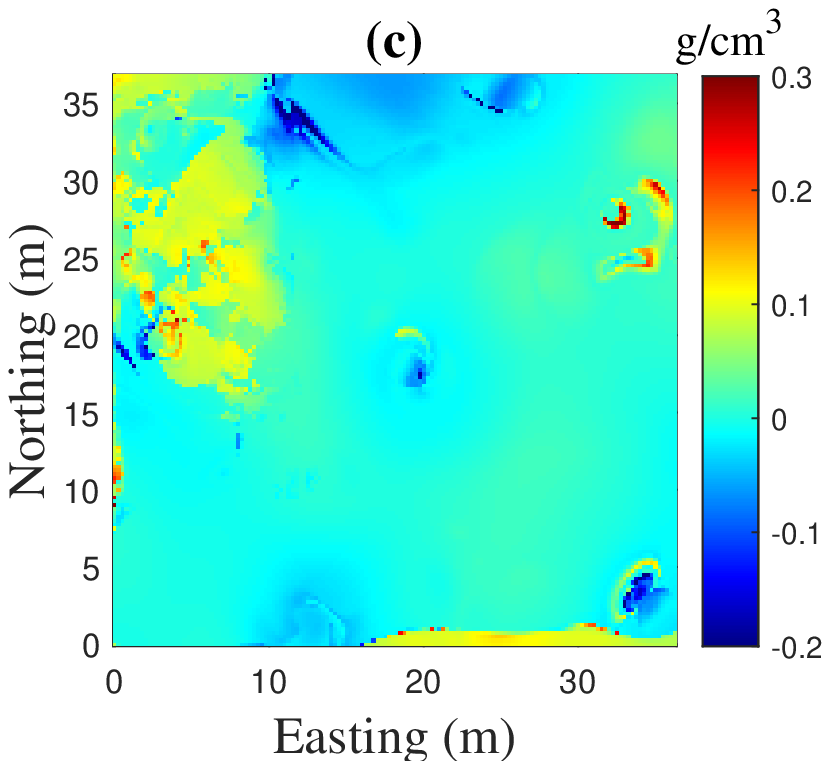}}
\caption{Depth sections of the reconstructed density model using the joint inversion algorithm applied on the data sets shown in~\cref{26a,26b} with Lagrange parameters, $\lambda^{(1)}=2$, $\lambda^{(2)}=5$. The sections are at depth: in~\cref{28a} $1$~km; in~\cref{28b} $3$~km; and in~\cref{28c} $5$~km.} \label{fig28}
\end{figure*}

\begin{figure*}
\subfigure{\label{29a}\includegraphics[width=.48\textwidth]{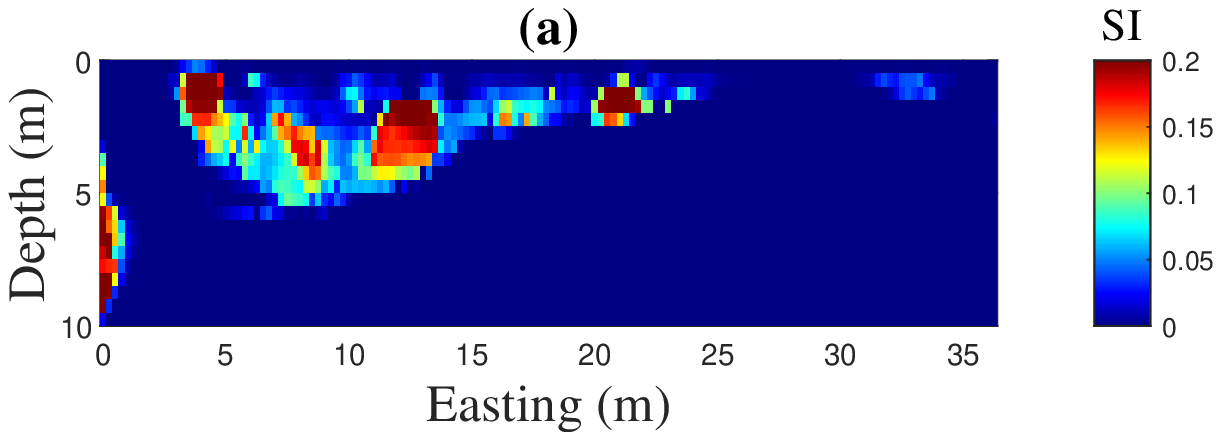}}
\subfigure{\label{29b}\includegraphics[width=.48\textwidth]{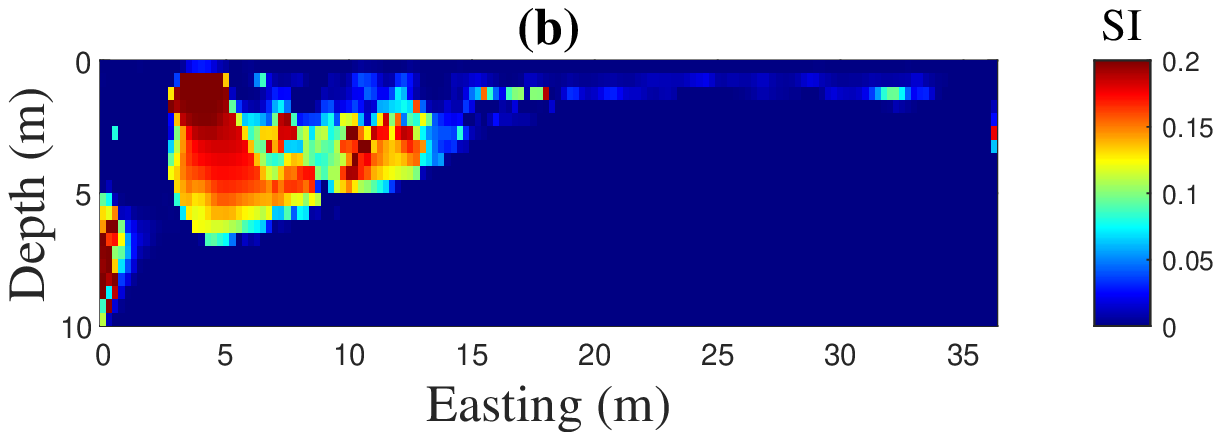}}
\subfigure{\label{29c}\includegraphics[width=.48\textwidth]{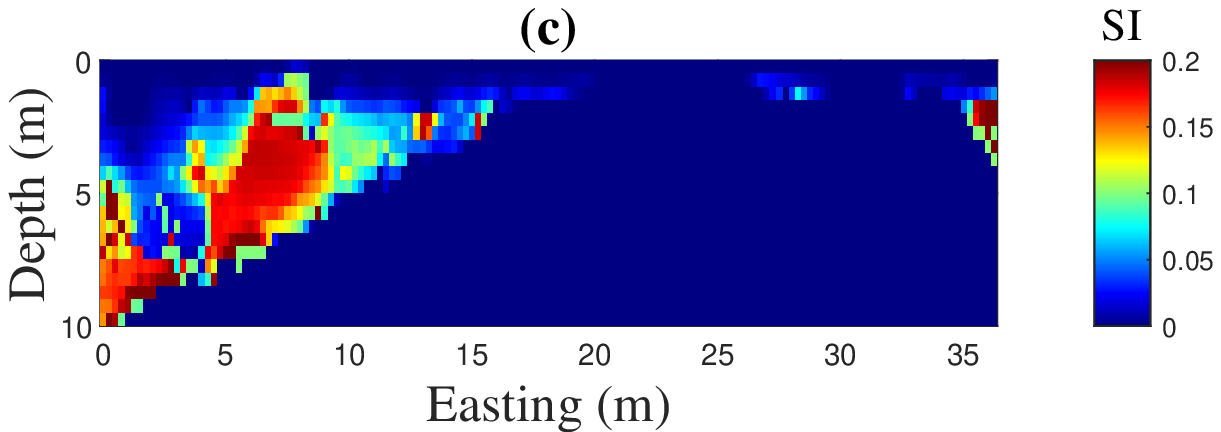}}
\caption{Cross-sections of the reconstructed magnetic susceptibility model using the joint inversion algorithm applied on the data sets shown in~\cref{26a,26b} with Lagrange parameters, $\lambda^{(1)}=2$, $\lambda^{(2)}=5$. The sections are at: in~\cref{29a} $16$~km Northing (over Pea Ridge); in~\cref{29b} $18$~km Northing; and in~\cref{29c} $30$~km Northing.} \label{fig29}
\end{figure*}

\begin{figure*}
\subfigure{\label{30a}\includegraphics[width=.44\textwidth]{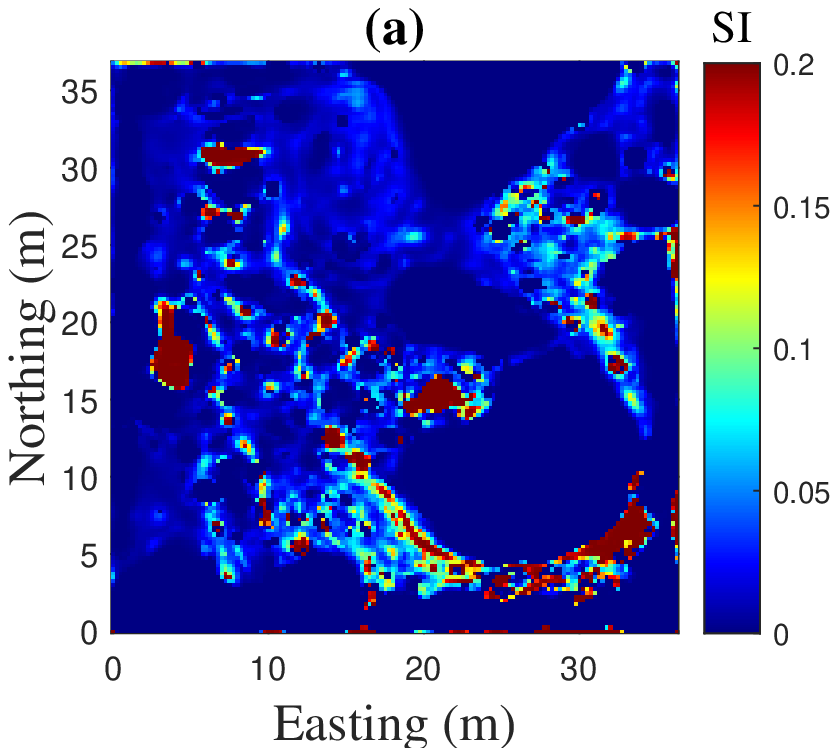}}
\subfigure{\label{30b}\includegraphics[width=.44\textwidth]{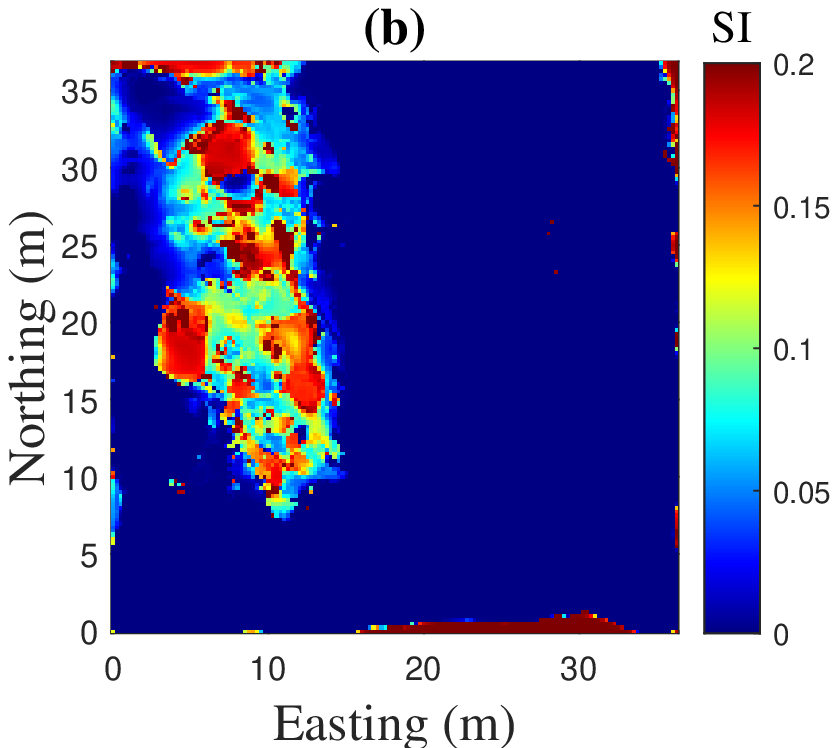}}
\subfigure{\label{30c}\includegraphics[width=.44\textwidth]{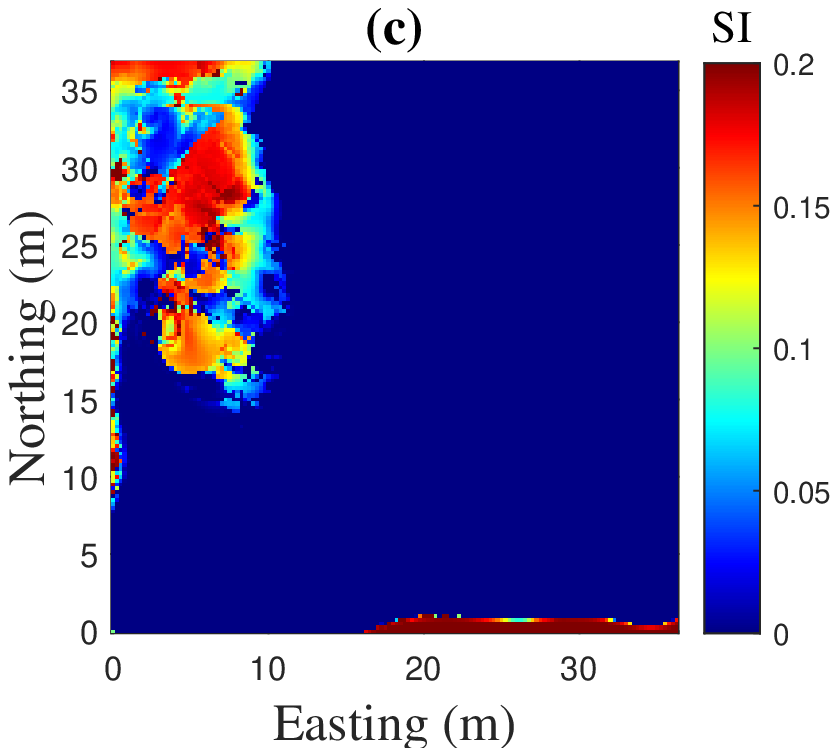}}
\caption{Depth sections of the reconstructed magnetic susceptibility model using the joint inversion algorithm applied on the data sets shown in~\cref{26a,26b} with Lagrange parameters, $\lambda^{(1)}=2$, $\lambda^{(2)}=5$. The sections are at depth: in~\cref{30a} $1$~km; in~\cref{30b} $3$~km; and in~\cref{30c} $5$~km.} \label{fig30}
\end{figure*}

\begin{figure*}
\includegraphics[width=.5\textwidth]{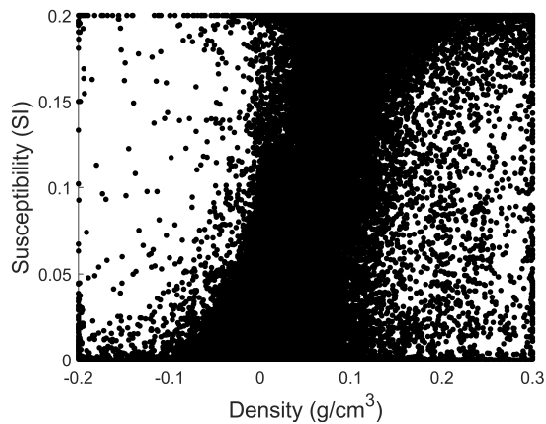}
\caption{The cross-correlation plot between the density and magnetic susceptibility model parameters for the models shown in~\cref{fig27,fig29}.} \label{fig31}
\end{figure*}

\begin{figure*}
\subfigure{\label{32a}\includegraphics[width=.45\textwidth]{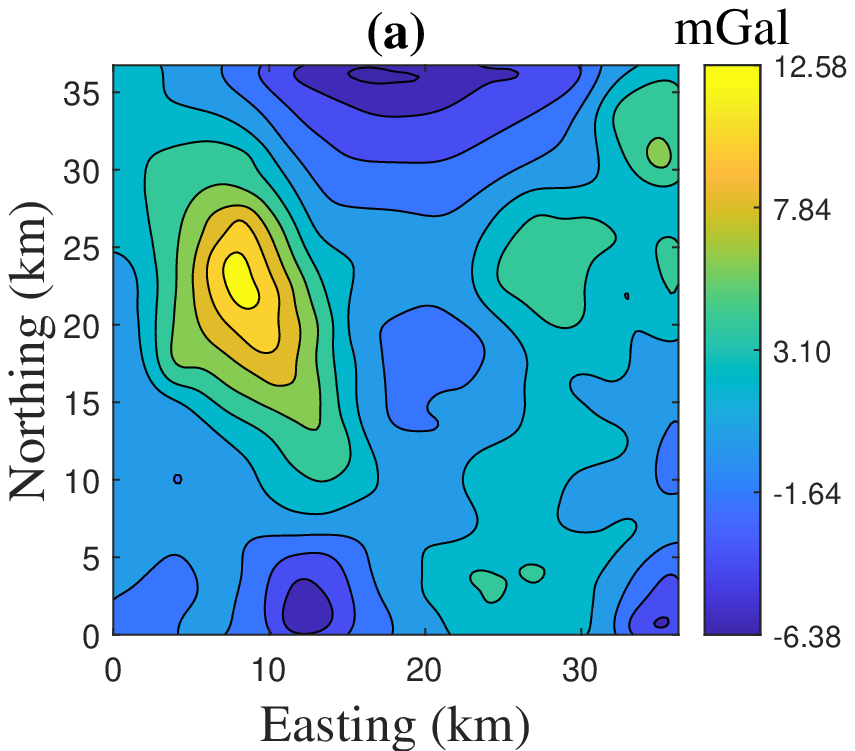}}
\subfigure{\label{32b}\includegraphics[width=.46\textwidth]{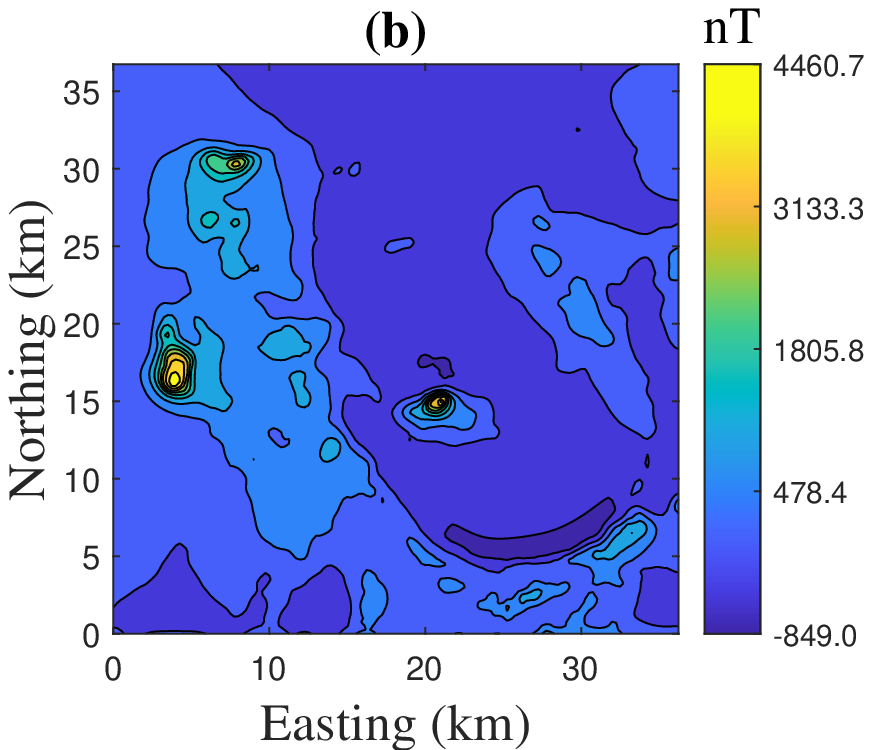}}
\caption{Data predicted by the models shown in~\cref{fig27,fig29}, obtained with  $\lambda^{(1)}=2$, $\lambda^{(2)}=5$. In~\cref{32a} the  vertical component of the gravity, and in~\cref{32b} the total magnetic field.} \label{fig32}
\end{figure*}

\section{Conclusions}\label{conclusion}
 An algorithm for the large-scale simultaneous joint inversion of gravity and magnetic data using the Gramian determinant coupling constraint has been presented. The objective function for the minimization is formulated in the  space of the weighted parameters, but the Gramian constraint is implemented in the original, unweighted, space of parameters. This approach provides good similarity between the reconstructed models, whereas the implementation of the Gramian in the space of weighted parameters yields a good correlation for the weighted parameters but not for the parameters in the original space, and the reconstructed results are, in consequence not similar.  
 
  The presented algorithm is efficient for both memory storage and access, as well as for computational cost, because of the use of the \bttb~structure of the sensitivity matrices which results when the data are measured on a uniform grid, with a consistently defined model volume. With this format, the dense sensitivity matrices are not stored, and all their operations (forward and transpose matrix vector multiplications) can be realized using the \twoDFFT. Moreover, the use of \rrcg~algorithm to minimize the objective function makes it advantageous to use the \bttb~structure of the matrices, and the resulting methodology provides an algorithm which is fast and effective for the joint inversion of gravity and magnetic potential field data.

Whereas a single independent inversion would generally rely on a single regularization parameter, requiring two parameters for inverting gravity and magnetic potential field data separately, the presented joint inversion approach relies on providing four parameters. Two of these correspond to the standard regularization parameters, but the two additional parameters are Lagrange parameters that weight the steepest ascent directions  of the coupling Gramian constraints for each parameter set. The results for simulated data show that the choice of Lagrange parameters is crucial for providing acceptable solutions which are well-correlated and geophysically meaningful.  While, the algorithm may converge quickly for small Lagrange parameters, the resulting parameter distributions may not be well-correlated. On the other hand, large parameters will provide well-correlated solutions but the impact of the  stabilizing terms is reduced and the algorithm does not converge quickly.  These observations suggest a strategy to chose an acceptable set of Lagrange parameters, specifically, the algorithm should be run with a small selection of parameters starting from small values, and gradually increasing the values to find the point at which a good correlation is first identified. This will provide a good set of parameters for finding an acceptable solution. 

The suggested joint inversion  algorithm uses the $L_1$-norm focusing stabilizer for the model parameters, which provides compact solutions with sharp boundaries. We demonstrated, however, that it is easy to modify the algorithm to  use a minimum length stabilizer, corresponding to the use of the $L_2$-norm stabilization, which may be sufficient for detecting subsurface targets for further investigation. The resulting targets are similar due to the use of the coupling constraint, but the resulting models are smooth and exhibit minimum structure. In this framework, the algorithm requires more iterations to converge, but the overall computational cost is reduced because the per iteration costs are substantially reduced. 

The developed joint inversion algorithm was applied on two synthetic examples. We showed that the algorithm could invert large-scale data sets very fast using a standard laptop computer. The structures of the obtained density and magnetic susceptibility models were approximately similar and were close to the original models. Finally,  we applied the algorithm on gravity and aeromagnetic data obtained over an untapped economic minerals area in northwest of Mesoproterozoic St. Francois Terrane, southeast of Missouri, USA. The reconstructed models illustrated that the igneous bodies in the area reside in the upper crust, extending to a maximum depth of 5 to 6 km, except for the north of the Bourbon mineral deposit in which the subsurface target is deeper. For the Pea Ridge deposit, the reconstructed models indicate a positive density and magnetic susceptibility body, that extends from $1$~km to approximately $2$~km, above a low density region. In the western section of the models, deeper high density and magnetic susceptibility bodies were obtained. Our results provided a valuable insight about the geometry of Mesoproterozoic volcanic bodies that are concealed beneath a thin Paleozoic sedimentary sequence. Developing the algorithm for application on other geophysical data sets is the subject of future work.

\begin{acknowledgments}
The work was supported by Geophysical Computational Imaging and Instrumentation Group for the project CCL2021HNFN0250 at Jilin University. Furthermore Rosemary Renaut acknowledges the
support of the 
 National Science Foundation (NSF) under grant  DMS-1913136. 

\end{acknowledgments}

\appendix
\section{Steepest ascent directions of the Gramian constraint}\label{steepest}
The Gramian constraint in the space of weighted parameters can be expressed as
\begin{align}\label{GramianA1}
\tilde{S}_G (\mt^{(1)},\mt^{(2)})=\left\vert   
\begin{array}{cc} \| \mt^{(1)} \|_2^2&(\mt^{(1)},\mt^{(2)}) \\ (\mt^{(2)},\mt^{(1)})& \| \mt^{(2)} \|_2^2 \end{array}  \right\vert, 
\end{align}
where here we introduce $\tilde{S}_G$ as distinct from $S_G$ calculated in the space of unweighted parameters. Now by \cref{weighting}, the weighted parameters $\mt^{(i)}$ are related to their unweighted counterparts by $\mt^{(i)} = W^{(i)}\bfm^{(i)}$, where $W^{(i)}$ is a diagonal weighting matrix, which is in general not a scalar multiple of the identity. In particular, even when $\Wh=\WL=\bfeye$, $\Wdepth$ is diagonal with entries dependent on depth levels. Hence $\| \mt^{(i)} \|^2_2 = \| W^{(i)}\bfm^{(i)} \|^2_2$, (also known as the  weighted norm $\| \bfm^{(i)} \|^2_{W^{(i)}}$), and $(\mt^{(1)},\mt^{(2)})= (W^{(1)}\bfm^{(1)}, W^{(2)}\bfm^{(2)})$, and it is immediate that $(\mt^{(1)},\mt^{(2)}) = (\mt^{(2)},\mt^{(1)})$ because all components are real. Therefore we have 
\begin{align}\label{GramianA2}
\tilde{S}_G &=\| \mt^{(1)} \|_2^2 ~ \| \mt^{(2)} \|_2^2 -(\mt^{(1)},\mt^{(2)})^2 \\ \label{Gramianrelb}
&=\| W^{(1)}\bfm^{(1)} \|_2^2 ~ \| W^{(2)}\bfm^{(2)} \|_2^2 -(W^{(1)}\bfm^{(1)}, W^{(2)}\bfm^{(2)})^2, 
\end{align}
and it is immediate that in general we do not have $\tilde{S}_G = S_G$. On the other hand, if $W^{(i)}=w^{(i)}\bfeye$, $i=1$, $2$, for some scalar $w^{(i)}$, we obtain $\tilde{S}_G=(w^{(1)}w^{(2)})^2 S_G$. 

We proceed now under the general case that the two Gramians are not the same, with diagonal weighting matrix $W^{(i)}$ not a scalar multiple of the identity. For  a weighted vector,  $\mt=D\bfm$ for diagonal matrix $D$, and with a general vector $\bfy$, we have 
\begin{align}\label{WeightGrad}
\nabla_{\mt} (\|\mt\|^2 )&= 2 \mt = 2 D \bfm \text{ and }  \nabla_{\mt} ((\mt, \bfy)^2) = 2  (\mt, \bfy) \bfy.
\end{align}
Hence,  the steepest ascent direction for $\tilde{S}_G$ to be used in \cref{FirstVariation} is given by 
\begin{align}\label{descentrela}
    \tilde{\bfl}_G^{(1)}= \nabla_{\mt^{(1)}} \tilde{S}_G&= 2 \|\mt^{(2)}\|^2 \mt^{(1)} - 2  (\mt^{(1)}, \mt^{(2)}) \mt^{(2)},\\
    & = 2 \|\bfm^{(2)}\|_{W^{(2)}}^2 (W^{(1)}\bfm^{(1)}) -2  (W^{(1)}\bfm^{(1)}, W^{(2)}\bfm^{(2)}) (W^{(2)}\bfm^{(2)}),\label{descentrelb}
\end{align}
with the roles interchanged for $\tilde{\bfl}_G^{(2)}$, and  we do not have a diagonal weighting $\tilde{\bfl}_G^{(i)} =  D^{(i)}{\bfl}_G^{(i)}$, for some diagonal matrices $D^{(i)}$. This explains, as seen in the presented results, that minimizing for the coupling in the weighted domain indeed gives a result which is correlated in the weighted domain. Further, we would not expect a result that is similar to the result obtained in the original domain; the search directions cannot be formulated as a diagonal weighting between the two approaches.  Equivalently, we cannot have gradients of $S_G$ and $\tilde{S}_G$ small, or zero,  for a common minimizer because there is no component wise direct proportionality between the ascent directions.

Suppose, on the other hand, that we use $S_G$ rather than $\tilde{S}_G$, based on using the parameters for the coupling in the original space, and the objective function is replaced by 
\begin{multline}\label{GlobalFunctionMixed}
P^{(\alpha,\lambda)}(\mt^{(1)},\mt^{(2)})=\sum_{i=1}^2 {\| \At^{(i)} \mt^{(i)}-\dotilde^{(i)} \|_2^2} +  \sum_{i=1}^2 { \alpha^{(i)} \| \mt^{(i)}-\mat^{(i)} \|_2^2}  + \\ {\lambda}~{S}_G (\bfm^{(1)},\bfm^{(2)}). 
\end{multline}
Then the ascent directions require $\nabla_{\mt^{(i)}} {S}_G$. Now we note that $\bfm^{(i)}=(W^{(i)})^{-1} \mt^{(i)}$, and as in~\cref{WeightGrad}, with now $D=(W^{(i)})^{-1}$,
\begin{align}\label{invWeightGrad}
\nabla_{\mt} (\|\bfm\|^2 )&= \nabla_{\mt} (\|D \mt\|^2 ) = 2 D^2  \mt = 2 D\bfm 
\text{ and }  \nabla_{\mt} ((\bfm, \bfy)^2) = 2 D (\bfm, \bfy) \bfy.
\end{align}
Hence we obtain 
\begin{align}\label{descentrelaS}
 \nabla_{\mt^{(1)}} {S}_G&= 2(W^{(1)})^{-1} \|\bfm^{(2)}\|^2 \bfm^{(1)} - 2 ( W^{(1)})^{-1} (\bfm^{(1)}, \bfm^{(2)})  \bfm^{(2)} =(W^{(1)})^{-1} \bfl_G^{(1)}\\
 \label{descentrelbS}
 \nabla_{\mt^{(2)}} {S}_G&= 2(W^{(2)})^{-1} \|\bfm^{(1)}\|^2 \bfm^{(2)} - 2 ( W^{(2)})^{-1} (\bfm^{(2)}, \bfm^{(1)})  \bfm^{(1)} =(W^{(2)})^{-1} \bfl_G^{(2)}.
\end{align}
Immediately, the ascent directions with respect to the weighted variables are  not given by a scalar multiple of the directions for the unweighted variables when switching between   $\tilde{S}_G$ and $S_G$, but they are component wise proportional to $\bfl_G$.  Indeed,  since the directions are weighted by $(W^{(i)})^{-1}$, this corresponds to using a different Lagrange parameter for each prism $j$ of the parameter domain. Furthermore, since $W^{(i)}$ depends on $\WL^{(i)}$ the parameter weighting changes with the iteration according to 
\begin{align}\label{entriesofWLinv}
(\WL^{(i)})_{jj}^{-1}=((\bfm^{(i)}-\bfma^{(i)})^2+\epsilon^2)_{jj}^{1/4} \ge \epsilon^{1/2}=3.1e-5.
\end{align}
Now, since in the implementation we use $\lambda^{(i)}$ fixed across all iterations, implicitly the algorithm  actually  adjusts the search directions by an iteratively changing set of Lagrange weights. We see that this means that for components with $\bfm_j^{(i)}-(\bfma)_j^{(i)}\approx 0$, the weighted Lagrangian parameter is very small, and little coupling is imposed for that component. On the other hand, when $\bfm_j^{(i)}-(\bfma)_j^{(i)} \gg 0$, $(\WL^{(i)})_{jj}^{-1}$ is dominated by 
$|\bfm_j^{(i)}-(\bfma)_j^{(i)}|$ and the coupling constraint is larger, but still small at convergence. Hence, if we ignore the weighting of the ascent directions, as we do in~\cref{RRCG:2,RRCG:3}, and use instead a single Lagrange parameter, $\lambda^{(i)}$, we would expect that this parameter is small, and smaller than if we used the  ascent directions for $\tilde{S}_G$, in~\cref{descentrela}, where there is no  weighting matrix, and in particular no weighting matrix $(W^{(i)})^{-1}$ which we would expect to have small entries. Contrary to the discussion for the ascent directions in the weighted domain, here we use component wise scaled directions for the original unweighted domain, and hence force the gradients for the original domain to be small or close to zero. Equivalently, the minimizer is at a point of high correlation in the original domain, as is observed in the results.

\end{document}